\documentclass[final,3p,authoryear]{elsarticle}

\usepackage[T1]{fontenc}
\usepackage{amsmath}
\usepackage{newtxtext}
\usepackage{newtxmath}
\usepackage{graphicx}
\usepackage{booktabs}
\usepackage{longtable}
\usepackage{array}
\usepackage{tabularx}
\usepackage{float}
\usepackage{placeins}
\usepackage{etoolbox}
\usepackage[hidelinks]{hyperref}
\usepackage{xurl}
\AtBeginEnvironment{longtable}{\footnotesize}
\newcommand{\pandocbounded}[1]{\resizebox{\linewidth}{!}{#1}}

\begin{document}

\begin{frontmatter}

\title{AIDA: An AI Outage Registry with Evidence from On-Chain MEV}

\author{Wenbin Wu}
\nonumnote{\textit{Email address:\space}\texttt{w.wu@jbs.cam.ac.uk}}
\address{Cambridge Centre for Alternative Finance, Cambridge Judge Business School, University of Cambridge\\25 Trumpington Street, Cambridge CB2 1QA, United Kingdom}

\begin{abstract}
Generative AI coding tools have lowered the cost of writing trading software, but no systematic dataset records when these services fail. We introduce the AI Incident and Disruption Archive (AIDA), a multi-vendor registry of 5,868 deduplicated outage incidents across 30 providers from 2025 to 2026, with foundation-model incidents verified against primary status-page or postmortem URLs. Pairing AIDA with 5,041,034 verified maximal extractable value (MEV) transactions by 8,516 Ethereum addresses, their contract creations, and a 22-month Base Layer-2 panel of 1,099,593,191 swaps and 29,120 private executors, we establish two findings. First, entry into MEV expanded where code and public state suffice, as active arbitrage addresses quadrupled from late 2024 to late 2026 and liquidators grew by a third, while sandwich attackers, who need private order flow and builder access, declined. Execution nonetheless remained concentrated, with the top fifty addresses executing two thirds of verified trades. Second, AI outages stall new trading code while deployed bots keep trading. On Ethereum, addresses releasing a private executor fall to 0.67 on the day of an eight-hour disruption (0.68 for executor creations during OpenAI outage hours) and to 0.66 at lag two before rebounding on day three; on Base, operator top-level creations fall to 0.76 during OpenAI hours and executor-deploying operators to 0.88 at lag two. Non-trading creations, automated factory clones, and live arbitrage and sandwich execution across both chains remain unchanged. Cheaper code thus lowers the barrier to building MEV contracts, while private order flow, latency and capital still determine who wins execution.
\end{abstract}


\end{frontmatter}

\section{Introduction}\label{introduction}

Artificial intelligence enters financial markets both as an engineering input that helps developers write and test software and as a potential run-time input that processes information while trading is under way \citep{Abis2024Knowledge, Korinek2023GenAI}. Generative assistants raise completed software-developer tasks by about a quarter \citep{Cui2026Developers, Peng2023Copilot} and speed up professional writing \citep{Noy2023Science}, customer-support agents who use generative assistants for several months retain their faster pace even when the assistant goes down, suggesting that the productivity gain partly embeds in learned routines rather than requiring continuous AI access \citep{Brynjolfsson2025Work}, and the gains from AI investment accrue largely to firms that were already large \citep{Babina2024AI}. These findings predict that cheaper software production widens the pool of developers who can build trading programs, while execution advantages that software alone does not confer, such as private order flow, latency and capital \citep{Budish2015HFT, Biais2015HFT}, keep market outcomes concentrated.

Public blockchains offer a unique setting to test both predictions at once, because they record software deployment and trade execution for the same participants on the same ledger. On Ethereum and Layer-2 rollups such as Base, automated searchers compete to capture maximal extractable value (MEV) through atomic arbitrage across decentralized exchange (DEX) pools, sandwich attacks around pending user swaps, and collateral liquidations in lending protocols \citep{Daian2020FlashBoys, Qin2022Quantifying, Zhou2021HFT, Qin2021Liquidations}. Each executed strategy is a transaction with an identifiable sender, and each new executor contract is a top-level or factory contract creation. Staying competitive requires continuous off-chain engineering as protocols upgrade and pools multiply, and multi-agent language-model harnesses now synthesize, test and refine MEV strategies against forked chain state before deployment \citep{Wen2026EVAGE}; once bytecode is on-chain, however, bots monitor state and submit transactions autonomously. Observing both margins for the same addresses resolves an identification problem that price-and-volume event studies face: with aggregate data alone one cannot tell whether a disruption affects the building of trading software, the execution of live trades, or both. If AI outages delay these operators\textquotesingle{} executor releases, their development workflow depends on AI coding services, and the behavior of their live trades during the same outages then tests run-time dependence directly.

We ask two questions over the 22-month target period from 1 January 2025 to 12 October 2026, with historical support from 5 September 2024. First, where did entry expand, and did execution concentration change alongside it? The three strategies differ in what they require beyond code. Atomic arbitrage and liquidations act on public state that any program can read, whereas sandwich attacks require low-latency visibility of pending orders and placement of the attacker\textquotesingle s bundle with a block builder. If cheaper software is the binding constraint, entry should expand in the first two strategies but not the third, and execution should remain concentrated among operators who hold complementary advantages. Second, does automated trading depend on AI services at run time, or only during development? When AI foundation-model providers experience service disruptions, do the addresses that run these strategies keep releasing code and executing trades?

Answering these questions requires transaction-level strategy identification and a systematic record of AI service disruptions. We identify atomic arbitrages, sandwich attacks and lending liquidations from primary-source rules over raw event logs, favoring precision over recall, across the full Ethereum span (5,041,034 verified strategy transactions by 8,516 addresses; Table 1) and a 22-month Base panel. To measure AI service availability we construct the AI Incident and Disruption Archive (AIDA), a multi-vendor registry of 5,868 deduplicated incidents across 30 providers and 17 foundation-model vendors (709 developer-relevant disruptions of 60 minutes to 24 hours; Figure 1). We classify each created contract by what it later does on-chain and estimate daily and hourly deployment models alongside matched episode-window comparisons of live execution, testing robustness across five operator and bot definitions from the MEV literature (Sections 2 and 3).

\begin{figure}
\centering
\pandocbounded{\includegraphics[keepaspectratio,alt={The AI Incident and Disruption Archive (AIDA) across proprietary foundation-model vendors and calendar months, January 2025 to October 2026. Panel A compares raw status reports (5,291), complete-linkage deduplicated incidents (3,085) and developer-relevant disruptions lasting 60 minutes to 24 hours (709 incidents, 2,586.5 exposure hours) across the 17 foundation-model vendors. Panel B plots monthly developer disruption hours across all 17 vendors, separating OpenAI and Anthropic from the remaining 15 international and Chinese foundation-model vendors.}]{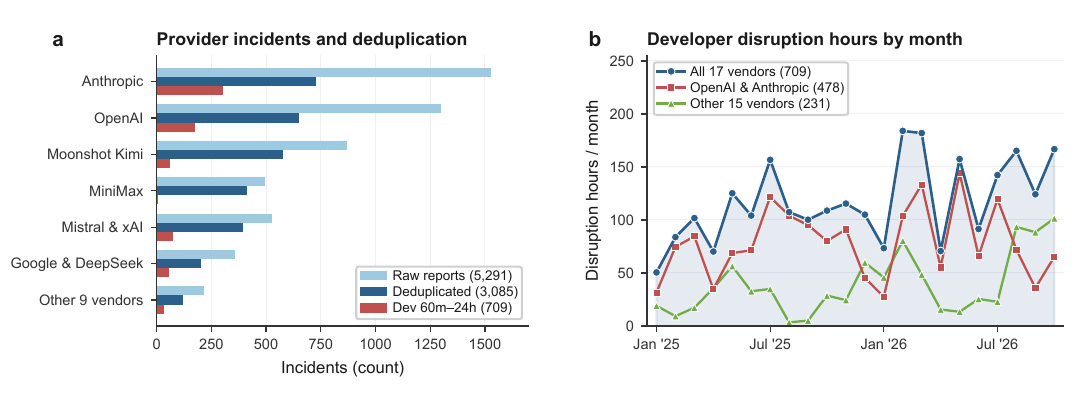}}
\caption{The AI Incident and Disruption Archive (AIDA) across proprietary foundation-model vendors and calendar months, January 2025 to October 2026. Panel A compares raw status reports (5,291), complete-linkage deduplicated incidents (3,085) and developer-relevant disruptions lasting 60 minutes to 24 hours (709 incidents, 2,586.5 exposure hours) across the 17 foundation-model vendors. Panel B plots monthly developer disruption hours across all 17 vendors, separating OpenAI and Anthropic from the remaining 15 international and Chinese foundation-model vendors.}
\end{figure}

Two findings emerge. First, entry into MEV expanded where code and public state suffice. Active arbitrage and liquidation addresses expanded markedly between the fourth quarter of 2024 and the third quarter of 2026, while sandwich attackers, who need private order flow and builder access, contracted (Table 2). The deployer base grew severalfold, yet execution remained concentrated among leading operators. Building and releasing trading contracts thus extends across a broad and growing base, while capturing execution depends on inputs that cheaper software does not provide.

Second, AI outages stall new trading code while deployed bots keep trading. On Ethereum, addresses releasing a new executor contract decline sharply on the day of a multi-hour disruption and rebound within days, consistent with a multi-day development pipeline (Table 3). On Base the same separation appears in operator-signed top-level creations and executor-deploying operators. Meanwhile, non-trading contract creations, automated factory clones, constructor-executed trades and live arbitrage and sandwich execution across both chains remain unchanged (Tables 3 and 4).

The two findings are complementary. The entry pattern shows that the expanding base of MEV developers enters exactly the strategies where code and public state suffice, consistent with cheaper software lowering the building barrier. The outage design then provides direct evidence that these developers rely on AI coding services, because when those services fail, new executor releases pause while the same operators\textquotesingle{} deployed bots trade on uninterrupted. Together, the findings establish that AI assists MEV in off-chain software engineering while live on-chain execution runs autonomously. Section 5 discusses how address structure, contract upgrades and venue coverage bound these measurements.

This study makes three contributions. First, we construct and release AIDA (\url{https://huggingface.co/datasets/venvoo/aida}), a multi-vendor AI outage registry extending cloud-outage forensics \citep{Gunawi2016Cloud, Acemoglu2012Network} to foundation-model infrastructure, where existing event studies have relied on one or two providers without root-cause categories \citep{Cheng2025Outages, Paydarzarnaghi2026Outages, Chang2026Democratization, Wang2026CryptoOutages, Cao2026CryptoLiquidity}. Second, pairing AIDA with on-chain deployments and trade execution lets us separate off-chain development from live execution, connecting evidence on task automation and AI-assisted knowledge production \citep{Acemoglu2018Automation, Abis2024Knowledge} to a market in which trading algorithms are continuously rewritten. Third, we add to the MEV and market-quality literature \citep{Aquilina2022Races, Makarov2020Crypto, Barbon2021Quality, Brolley2023DEX, Park2023AMM, Capponi2025AMM, Condie2025AMM, Muck2025Arbitrage, Lehar2025Uniswap} a transaction-level measurement of who builds trading contracts and who captures execution across strategies and chains.

\section{Automated trading on blockchains}\label{automated-trading-on-blockchains}

\subsection{Three strategies and their on-chain signatures}\label{three-strategies-and-their-on-chain-signatures}

Automated trading on programmable blockchains centers on atomic DEX arbitrage, sandwich attacks and lending-protocol liquidations. Each leaves a distinct event signature. Early measurement work classified contracts by the median number of pools they touched or by adjacent block positions, which conflates routing and victim flow with extraction; we apply transaction-level rules instead, each illustrated with an audited transaction in Figure 2.

Atomic arbitrage exploits transient price differences across pools within a single transaction. Following Section IV of \citet{Qin2022Quantifying} and the graph-cycle analysis of \citet{Chi2024Remeasuring}, an arbitrage cycle requires a closed loop of at least two swaps across at least two distinct pools within one transaction; ordering the legs by log index, each interior hop must feed the next hop\textquotesingle s input token with an integer amount no larger than the preceding hop\textquotesingle s output, and the closing hop must return at least the initial input. Panel A of Figure 2 shows a three-leg cycle across three Uniswap V4 pools, ETH to USDT to USDC and back to more ETH than was spent. All qualifying cycles in one transaction collapse into a single arbitrage record for the sender. Because the sequential rule insists on chronological order and integer-exact hops, it excludes some profitable multi-branch arbitrages, so our arbitrage counts are lower bounds.

Sandwich attacks extract value from a pending user swap by front-running it in the same block and pool and back-running it in the opposite direction after it executes \citep{Zhou2021HFT, Park2023AMM}. The attacker must see the victim\textquotesingle s order before inclusion, in the public mempool or through a private order-flow channel, and must place a bundle with a block builder so that the three legs land in order \citep{Lehar2023Battle, Weintraub2022Flashbots, Heimbach2023PBS}; the strategy depends on low-latency order visibility and builder access, two inputs that software alone does not provide. Our rule requires the front-run, victim and back-run in that order in one block and pool, the front-run and victim trading the same pair in the same direction and the back-run reversing it, the two attacker legs sharing a sender or an executor whose common control is independently established (a shared public router never suffices), and the back-run input within 90 to 110 percent of the front-run output in exact integer arithmetic. Each front-run maps to exactly one back-run; pairs that require a one-to-many or many-to-one mapping are recorded as candidates and excluded from counts. Panel B of Figure 2 shows such a pattern on a Uniswap V2 pool; activity is attributed to the two attacker transactions only, and the victim never enters cohorts or counts.

Collateral liquidations occur when a borrower\textquotesingle s health factor falls below the protocol threshold, allowing a liquidator to repay part of the debt for discounted collateral \citep{Qin2021Liquidations}. We decode \texttt{LiquidationCall} events in Aave V2 and V3 and \texttt{LiquidateBorrow} events in Compound V2, resolving protocol versions from the registry of canonical lending-market contract addresses because Aave V2 and V3 share one event signature, and we require strictly positive repaid-debt and seized-collateral integers. Panel C of Figure 2 shows an Aave V3 liquidation in which the sender calls its executor to repay USDC debt for a borrower and seize WBTC collateral; activity is attributed to the sender. Panel D shows the corresponding verification of a Raydium swap on Solana, routed through the Jupiter aggregator (Supplementary Section S3).

\begin{figure}
\centering
\pandocbounded{\includegraphics[keepaspectratio,alt={Transaction-level strategy identification and operator attribution across Ethereum and Solana. Panel A shows a verified three-leg Ethereum arbitrage cycle across three Uniswap V4 pools with a closing gross surplus of 21.07 basis points; Panel B shows signed WETH flows for a Uniswap V2 sandwich pattern, separating the attacker\textquotesingle s front-run and back-run from the excluded victim swap (hollow hatched bar). Panel C shows an Aave V3 liquidation attributed to the transaction sender while excluding the borrower (hollow node), and Panel D shows a verified Solana Raydium CLMM swap routed through Jupiter.}]{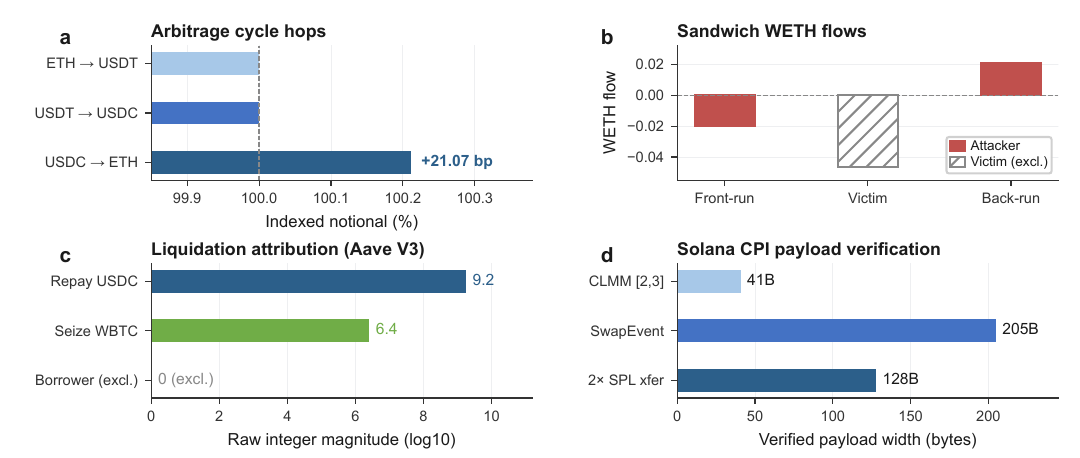}}
\caption{Transaction-level strategy identification and operator attribution across Ethereum and Solana. Panel A shows a verified three-leg Ethereum arbitrage cycle across three Uniswap V4 pools with a closing gross surplus of 21.07 basis points; Panel B shows signed WETH flows for a Uniswap V2 sandwich pattern, separating the attacker\textquotesingle s front-run and back-run from the excluded victim swap (hollow hatched bar). Panel C shows an Aave V3 liquidation attributed to the transaction sender while excluding the borrower (hollow node), and Panel D shows a verified Solana Raydium CLMM swap routed through Jupiter.}
\end{figure}

\subsection{Software deployment and trade execution for the same addresses}\label{software-deployment-and-trade-execution-for-the-same-addresses}

Automated trading combines off-chain engineering with on-chain deployment and execution. Searchers write, compile and test routing and liquidation logic against local forks of the chain \citep{Wen2026EVAGE}; when an operator deploys a new executor, the chain records a contract creation, either a top-level transaction with a null destination signed directly by the operator\textquotesingle s key or an internal creation performed by a factory inside a transaction that calls it. Off-chain tuning, mempool monitoring and order submission leave no deployment trace, and even among visible creations a single event can be a substantive revision or a routine redeployment. Creations and strategy transactions therefore capture two stages of the trading workflow once raw creations are separated by what the created contract does and how it is triggered. Some operators execute atomic trades inside a contract\textquotesingle s constructor without subsequent calls, so that the contract-creation transaction itself constitutes a trade execution, many deploy token, helper or test contracts unrelated to MEV, and on low-gas Layer-2 rollups such as Base existing contracts also spawn large numbers of internal factory clones automatically during routine trading. We therefore classify Ethereum creations by the created contract\textquotesingle s subsequent role into persistent private executors that go on to execute identified MEV transactions, broader swap-receiving trading contracts, single-use constructor-executed trades, and non-trading contracts, while on Base we also separate top-level creations signed directly by an operator key, distinct operators deploying an executor, and per-operator capped executor creations from uncapped internal factory spawns.

AI coding assistants and foundation models can enter this workflow during off-chain writing and testing or during live monitoring and parameter selection. On the live execution path, Ethereum produces a block at twelve-second intervals and Base at two-second intervals, and block builders or sequencers order transactions within milliseconds, so a bot that wins execution evaluates state and submits in milliseconds; a cloud language-model call that takes several seconds cannot sit inside that execution cycle, and any live role for AI is confined to background monitoring or parameter updates that the bot applies locally. The chain does not record API requests to AI providers, which operators subscribe to which vendor, or whether an operator switches to a substitute during an interruption. A release passes through writing, local fork testing and staging before deployment, so an interruption during final testing delays a same-day release, an interruption while writing or adapting routing logic delays a release scheduled one or two days later, and interrupted builds then finish and clear the pipeline in a rebound while existing monitoring and submission software keeps calling already-deployed contracts. We therefore track daily deployment counts over a four-day window alongside hourly within-day models and matched episode windows.

\subsection{Outcome definitions and competition accounting}\label{outcome-definitions-and-competition-accounting}

In the daily panel over days \(t \in \{1, \dots, T\}\), let \(Y_t\) denote the number of distinct population addresses that deploy on day \(t\) (or total creations, or creations capped at 3 per address-day) and let \(H_t\) denote the number of hours on day \(t\) during which a qualifying coding-service disruption was active. Following \citet{Hausman1984Count} and \citet{SilvaTenreyro2006PPML}, we estimate the Poisson pseudo-maximum-likelihood (PPML) specification

\[E[Y_t \mid H_{t}, \dots, H_{t-3}, d_t, m_t] = \exp\!\left(\alpha + \sum_{k=0}^{3} \beta_k H_{t-k} + \gamma_{d(t)} + \delta_{m(t)}\right),\]

where \(\gamma_{d(t)}\) and \(\delta_{m(t)}\) are day-of-week and calendar-month fixed effects and standard errors are clustered by Monday-anchored calendar week alongside a 28-day calendar-block bootstrap. For an eight-hour disruption (\(D = 8\)) on day 0 followed by three clean days, each lag \(k\) has rate ratio \(\exp(D\beta_k)\), and the four-day cumulative effect is computed directly from predicted counts across the panel,

\[\Delta_{0:3}(D) = \frac{\sum_{k=0}^{3} T^{-1} \sum_{t=1}^{T} \left(\hat{\mu}_t(H_{t-k}=D, H_{t-j\neq t-k}=0) - \hat{\mu}_t(0,0,0,0)\right)}{4 \cdot T^{-1} \sum_{t=1}^{T} \hat{\mu}_t(0,0,0,0)},\]

with delta-method week-cluster intervals and 28-day block-bootstrap percentile intervals. Alongside the daily panel, we estimate an hourly within-day PPML specification across the 15,600 UTC hours of the target period on both Ethereum and Base \citep{Correia2020PPMLHDFE} with UTC-date and hour-of-day fixed effects and standard errors clustered by UTC date, where \(\exp(\theta_g)\) measures the rate ratio for a fully covered outage hour of provider group \(g\) relative to non-outage hours net of date and clock-hour effects. Because the daily panel is estimated for executor releases and for the benchmark creation outcomes, we also report the ratio of the executor rate ratio to the benchmark rate ratio with a week-cluster interval as a direct test of whether trading software responds differently from the other contracts the same operators deploy. On the execution margin, \emph{strategy transactions} count deduplicated verified strategy transactions within a matched episode window, \emph{active operators} count distinct addresses with at least one verified transaction, and \emph{competitive share} is the smaller group\textquotesingle s fraction of verified transactions within the fixed cohort.

Competitive accounting for the liquidation-window analysis uses ranks fixed before each episode. Within an episode\textquotesingle s 90-day discovery window we rank eligible addresses by verified strategy transactions, breaking ties by address, and assign the first 50 addresses to the leading group and the remaining, individually less active addresses to the smaller group. Because membership and labels are frozen at UTC midnight 28 days before onset, the same addresses populate the two groups in the incident window and in the reference windows of that episode.

For an episode \(e\) with incident-window count \(Q_{eT}\) and at least 2 clean reference windows \(c \in \{1, \dots, m_e\}\), let \(\bar{Q}_{eC} = m_e^{-1} \sum_{c=1}^{m_e} Q_{ec}\) denote the expected reference count. The aggregate rate ratio across matched episodes is

\[R_Q = \frac{\sum_e Q_{eT}}{\sum_e \bar{Q}_{eC}}.\]

Averaging references within each episode before summing across episodes weights each episode by its expected activity scale instead of its number of surviving reference dates. Applying the same aggregator to active-operator counts \(A_{eT}\) and \(\bar{A}_{eC}\) gives the participation ratio \(R_A\) and the intensity ratio \(R_I\) (transactions per active-operator window), which satisfy

\[R_Q = R_A R_I, \qquad R_A = \frac{\sum_e A_{eT}}{\sum_e \bar{A}_{eC}}, \qquad R_I = \frac{\sum_e Q_{eT} / \sum_e A_{eT}}{\sum_e \bar{Q}_{eC} / \sum_e \bar{A}_{eC}}.\]

The identity is evaluated with paired bootstrap draws that resample \(Q\) and \(A\) together. The smaller group\textquotesingle s share is defined on episodes where both the incident total and the reference total are positive; we take the difference between the incident share and the pooled reference share for each defined episode and average the differences with episode duration as weights. Episodes with zero total liquidations remain in the \(R_Q\) sample and do not enter the share calculation, without zero imputation.

\section{Data and empirical design}\label{data-and-empirical-design}

\subsection{Observation period, block boundaries and verified coverage}\label{observation-period-block-boundaries-and-verified-coverage}

The target period runs from 1 January 2025 to 12 October 2026 UTC. To form complete 90-day cohorts ending 28 days before the earliest January 2025 episodes, historical support extends back to 5 September 2024, so the full ledger spans 768 days (Figure 3). Block boundaries are fixed from the timestamps of the produced block headers, so each day and episode window maps to an exact block interval.

The liquidation ledger, the contract-creation archives and the swap ledger all cover the full span with complete verified block coverage (Table 1). Deployment trends of verified MEV addresses are read from an archive of the successful transactions those addresses sent, in which top-level creations are the transactions with a null destination. For deployment around outages we widen the population to all addresses on the chain that run these strategies by combining a census of successful contract creations over the span from the public block archive and the internal-transaction records of private-executor factories, the destinations of identified MEV transactions (several thousand of which are private executors called by at most three senders), and the creation records of those private executors, so that a deployer is counted whether it created its executor directly, through a factory, or attached executable code directly to its own wallet under account abstraction. Combining the direct senders of verified MEV transactions with the several hundred separate addresses that deployed their private executors gives the baseline chain-wide MEV population of 8,997 addresses.

So that findings do not depend on a single MEV detection algorithm, four further operator definitions vary how bots and searchers are identified (Table 3). These cover deployers of single-caller exclusive trading contracts with at least ten swaps motivated by the proxy-contract bot architecture described by \citet{Daian2020FlashBoys}, deployers of private trading contracts called by at most three addresses, all deployers of non-pool, non-protocol contracts that execute a DEX swap, and direct MEV sender externally owned accounts (EOAs, standard key-controlled wallets) without separate deployer keys. Within each population, creations are typed by the created contract\textquotesingle s on-chain behavior. In the baseline population, a few thousand creations deploy private executor contracts that go on to execute verified MEV transactions, a larger number deploy other swap-receiving contracts, some thirteen thousand single-use creations by a few dozen senders execute an MEV trade inside their own constructor, and more than nine tenths of the persistent contract creations are non-trading contracts (Table 3). Constructor-executed trades are tracked separately as run-time execution, and the full set of persistent contract creations, dominated by non-trading contracts, serves as an in-population specificity benchmark.

The swap ledger covers the full block span (Supplementary Section S1.1). It selects transactions whose logs carry one of nine pool-level swap event signatures across Uniswap V2, V3 and V4, PancakeSwap V3, Algebra, Balancer V2 and two Curve exchange events, and retains all logs of each selected transaction so that cycle detection sees the whole transaction. Auditing one unfiltered day of Ethereum logs (15 June 2026) bounds what this protocol filter misses, showing that transactions swapping only on pools outside the list, chiefly Fluid DEX, Curve\textquotesingle s newer crypto pools, DODO, Balancer V1 and V3 and one further concentrated-liquidity design, amount to under two percent of the selected count, and that about four percent of selected transactions carry a leg on such a pool that the decoder does not read, so a cycle through it is counted as broken. Request-for-quote fills and bridges are not public pool swaps and fall outside the arbitrage definition (Supplementary Section S1.2).

\begin{figure}
\centering
\pandocbounded{\includegraphics[keepaspectratio,alt={Study setting, observation coverage, monthly matched episodes and prospective cohort timeline. Panel A compares verified observation coverage across the historical support period and the target period for Ethereum liquidations, creations and swaps alongside the bounded Solana strata. Panel B plots monthly matched disruption episodes in the extended full-period sample, the baseline cloud-screened sample and the balanced sample matched on both calendars, and Panel C shows the 90-day cohort discovery window frozen 28 days before episode onset.}]{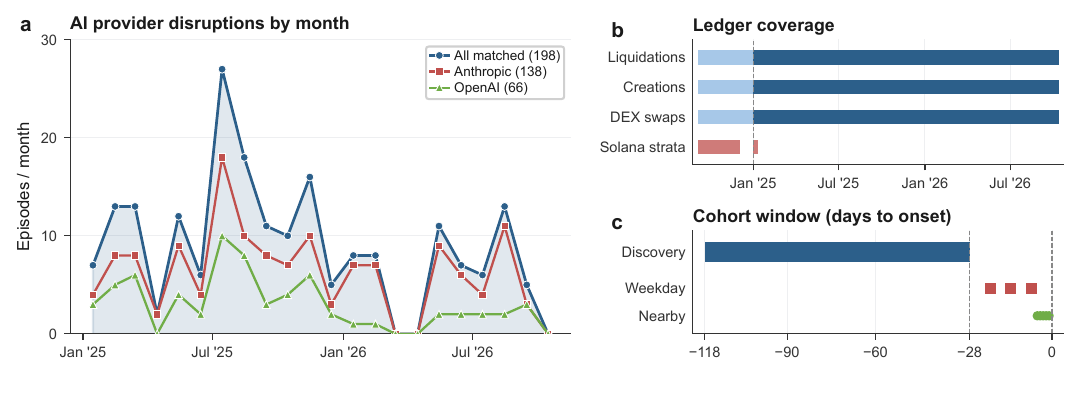}}
\caption{Study setting, observation coverage, monthly matched episodes and prospective cohort timeline. Panel A compares verified observation coverage across the historical support period and the target period for Ethereum liquidations, creations and swaps alongside the bounded Solana strata. Panel B plots monthly matched disruption episodes in the extended full-period sample, the baseline cloud-screened sample and the balanced sample matched on both calendars, and Panel C shows the 90-day cohort discovery window frozen 28 days before episode onset.}
\end{figure}

\subsection{The AI Incident and Disruption Archive, reference calendars and cloud screening}\label{the-ai-incident-and-disruption-archive-reference-calendars-and-cloud-screening}

Prior financial and economic outage studies draw on single-provider status lists without separating developer surfaces from consumer web interfaces or recording root causes. We construct the AI Incident and Disruption Archive (AIDA), a multi-vendor registry covering January 2025 through October 2026 across 30 providers organized into four layers: proprietary foundation-model APIs, managed cloud AI hosts, coding environments and developer platforms, and cloud, package and RPC infrastructure controls (Figure 1, Table 1 and Supplementary Section S3.2). We ingest 8,189 raw incident reports from official Statuspage.io and incident.io JSON endpoints, vendor engineering postmortems, Wayback Machine snapshots, StatusGator and IsDown mirrors, and news or developer-forum reports, and apply complete-linkage temporal and component deduplication to collapse split or hourly re-filed tickets (such as 35 consecutive hourly tickets opened by Moonshot Kimi during a single 10 July 2026 outage) into 5,868 distinct incidents, 1,543 corroborated by multiple sources. Fields are resolved along a source-authority hierarchy (official status pages and incident feeds first, followed by archive snapshots, status aggregators and community reports) with onset anchored to the earliest reported start of disruption, whether a vendor\textquotesingle s initial investigation update, an identified impact-start time or the onset time recorded in a retrospective postmortem. Primary-source verification against official status JSONs and archive captures confirms 2,977 incidents overall and 94.8 percent of foundation-model incidents.

Restricting to the 17 providers that train and serve proprietary foundation models (Anthropic, OpenAI, Google, xAI, Mistral, Cohere, Perplexity, Meta, and nine Chinese platforms led by Moonshot Kimi, MiniMax, DeepSeek, Zhipu, Doubao, Baidu, Qwen and Hunyuan) and retaining developer-relevant API, SDK, coding-agent and model-inference disruptions lasting 60 minutes to 24 hours yields 709 foundation-model disruptions totaling 2,586.5 exposure hours (668 inside the 15,600-hour target span, covering 2,215 non-overlapping outage hours), alongside 2,710 cloud, developer and RPC infrastructure incidents across 10 control providers (Cloudflare, QuickNode, Alchemy, GitHub, Infura, AWS, GCP, Azure, npm and PyPI). Among foundation-model developer disruptions with a stated cause, upstream cloud dependencies account for 27 percent, deployment or configuration changes 20 percent, capacity overload 19 percent, software bugs 15 percent and network routing 14 percent, while 95 percent of vendor status reports omit root causes. For the baseline Anthropic and OpenAI episode-window and daily dose-response panels, merging overlapping or adjacent developer-relevant intervals and dropping 18 multi-day administrative and compliance-API degradations longer than 24 hours yields 404 merged episodes, most falling inside the target window (Figure 3, Panel B).

For the matched-episode design, each episode is evaluated over its exact half-open UTC interval and paired with same-duration, same-clock intervals on nearby days (one to 5 days before onset) and on the same weekday one, two and three weeks earlier. A reference window that overlaps another provider incident or a public cloud disruption is removed, and an episode is matched on a calendar only if at least 2 clean references survive. Because changing the calendar changes both the comparison dates and the set of eligible episodes, we report each calendar across all of its matched episodes and in the balanced sample of episodes matched on both calendars.

The public cloud-status records used for screening are audited against primary sources. Four regional cloud-status records whose incidents began in late April 2026 lack verified closing timestamps; treating those open-ended intervals as active cloud disruptions screens out subsequent candidate windows and leaves episodes after late April 2026 without the required clean references. The baseline cloud-screened sample therefore retains 156 nearby-day liquidation episodes through late April 2026, while the extended full-period sample retains all 198 matched episodes through October 2026 without screening on the four unclosed regional records (Supplementary Section S6). Table 4 reports the episode count for each specification. On the episodes present in both samples the estimates are identical, so differences between the samples are composition effects.

\subsection{Inference, precision and reproducibility}\label{inference-precision-and-reproducibility}

For the episode-window estimates, confidence intervals come from 4,000 paired block-bootstrap draws that resample Monday-anchored UTC calendar weeks of episode onsets \citep{Cameron2008Cluster}, with 14-day blocks reported across the diagnostic cases in Supplementary Table S2. Weekly clusters preserve within-week dependence and the joint covariance of transaction, operator and group counts; they do not remove all dependence from reference windows that two episodes in different weeks share, and the fortnight sensitivity narrows that gap without closing it. An interval is reported only when at least 95 percent of draws have a positive reference denominator. For the daily panel, standard errors are clustered by Monday-anchored calendar week and checked against a 28-day calendar-block bootstrap. The AIDA dataset (\url{https://huggingface.co/datasets/venvoo/aida}) and replication package (\url{https://github.com/dthinkr/agentic-mev}) reconstruct the estimates, diagnostics and figures directly from the transaction-level datasets.

\subsection{Cross-chain extensions and quota announcements}\label{cross-chain-extensions-and-quota-announcements}

Three complementary analyses extend beyond the Ethereum outage design. First, to test whether the separation between off-chain development and on-chain execution holds on a high-throughput Layer-2 rollup where gas costs are orders of magnitude lower, we assemble a full 22-month panel on Base over the same 15,600 UTC hours from 1 January 2025 to 12 October 2026 (blocks 24,449,000 to 52,530,127), capturing 1,099,593,191 DEX pool-swap transactions, 29,120 private executor contracts (16,195 arbitrage, 50 sandwich and 12,875 other) across 21,026 caller addresses, 21,880 executor creations with observed creation traces, and 93,694 top-level contract creations signed directly by 2,888 MEV operators (Supplementary Section S3.2). Second, on Solana, where program upgrades replace code in place and routers invoke concentrated-liquidity pools through cross-program invocations, we verify loader and swap decoding on bounded historical strata and establish the archive volume that a full cohort panel would require (Supplementary Section S3.1). Third, around six public AI quota and access announcements in 2026 we track daily deployments by frozen liquidation cohorts from seven days before to fourteen days after each announcement (Supplementary Section S4).

\section{Results}\label{results}

\subsection{Verified evidence and validation}\label{verified-evidence-and-validation}

Across the swap ledger the decoder resolves nearly nine tenths of queried swap logs, and the ledger holds roughly two and a half million arbitrage transactions and a similar number of sandwich attacker transactions (Table 1). Together with the verified liquidations, the baseline sample comprises 5,041,034 distinct strategy transactions by 8,516 addresses. Table 1 also separates the historical support and target segments, reports the graph-cycle and multi-victim candidates that the sequential baseline excludes, and places the chain-wide creation census beside the strategy ledger; Supplementary Section S1.1 reports coverage by block interval.

\begin{table}[htbp]
\centering
\caption{The AIDA multi-vendor AI outage registry and verified on-chain MEV evidence.}
\label{tab_measurement_census}
\scriptsize
\renewcommand{\arraystretch}{1.10}
\setlength{\tabcolsep}{2.5pt}
\begin{tabular*}{\linewidth}{@{\extracolsep{\fill}}lrrrr@{}}
\toprule
Provider group or chain segment & Span & Raw / blocks & Verified / arb txs & Dev 60m-24h / sand \& liq \\
\midrule
\multicolumn{5}{@{}l@{}}{\textbf{Panel A: The AI Incident and Disruption Archive (AIDA, 30 providers)}} \\[1pt]
Anthropic (Claude API, Console, Claude Code) & 2025Q1-2026Q4 & 1,529 reports & 727 (707 verified) & 303 (838.2h) \\
OpenAI (API, Codex, Responses, Batch) & 2025Q1-2026Q4 & 1,299 reports & 650 (628 verified) & 175 (887.0h) \\
Chinese foundation-model vendors (9 providers) & 2025Q1-2026Q4 & 1,638 reports & 1,151 (1,088 verified) & 96 (286.0h) \\
Other international model vendors (6 providers) & 2025Q1-2026Q4 & 825 reports & 557 (500 verified) & 135 (575.3h) \\
All 17 proprietary foundation-model vendors & 22 months & 5,291 reports & 3,085 (2,923 verified) & 709 (2,586.5h) \\
Cloud AI, coding tools \& 10 cloud/RPC infra controls & 22 months & 2,898 reports & 2,783 (54 verified) & 2,710 infra controls \\
\textbf{Total AIDA registry (30 providers, deduplicated)} & \textbf{Jan 2025-Oct 2026} & \textbf{8,189 reports} & \textbf{5,868 (2,977 verified)} & \textbf{709 dev (2,215h)} \\
\addlinespace[3pt]
\multicolumn{5}{@{}l@{}}{\textbf{Panel B: On-chain MEV and contract creation panels (Ethereum L1 and Base L2)}} \\[1pt]
Ethereum swaps, historical support & 2024Q3-Q4 & 845,228 & 339,682 & 670,361 \\
Ethereum swaps, target period & 2025Q1-2026Q4 & 4,653,951 & 2,247,937 & 1,888,276 \\
Ethereum swaps, full baseline sample & 768 days & 5,499,179 & 2,587,619 & 2,558,637 \\
Ethereum swaps, graph-cycle \& multi-victim additions & 768 days & 5,499,179 & 348,284 & 401,045 \\
Ethereum liquidations, decoded markets & 768 days & 5,499,179 & 0 & 38,225 \\
Ethereum creations (8,997 operators; 96 factories) & 768 days & 5,499,179 & 2,892 & 139,146 \\
Base L2 22m panel (21,026 operators; 93,694 signed creations) & Jan 2025-Oct 2026 & 28,081,127 & 16,195 (26,165,900 txs) & 50 (1,099,593,191 swaps) \\
\bottomrule
\end{tabular*}
\vspace{3pt}
\begin{minipage}{\linewidth}
\footnotesize Notes: AIDA spans 1 Jan 2025 to 28 Oct 2026 UTC across 30 providers (2,215 non-overlapping outage hours inside the 15,600-hour on-chain target span). Ethereum blocks {[}20680663, 26179842) (5 Sep 2024 to 12 Oct 2026); creation columns report trading executors (2,892) and all persistent contracts (139,146). Base blocks {[}24449000, 52530127) (1 Jan 2025 to 12 Oct 2026).
\end{minipage}
\end{table}

A few dozen transactions whose token graphs are too dense to search exhaustively under the 100,000-state cycle cap are recorded as unresolved, receiving neither a positive nor a negative arbitrage label (Supplementary Section S5).

\subsection{Entry and concentration across strategies, 2024 to 2026}\label{entry-and-concentration-across-strategies-2024-to-2026}

Table 2 compares participation and concentration across the three strategies between the fourth quarter of 2024 and the third quarter of 2026; Supplementary Section S1.3 reports the full quarterly series. Nearly all verified addresses execute a single strategy, and only a handful execute all three.

\begin{table}[htbp]
\centering
\caption{Participation and concentration by strategy and period.}
\label{tab_entry_concentration}
\scriptsize
\renewcommand{\arraystretch}{1.10}
\setlength{\tabcolsep}{2.5pt}
\begin{tabular*}{\linewidth}{@{\extracolsep{\fill}}llrrrr@{}}
\toprule
Strategy & Period & Verified txs & Active addresses & Top-50 share & Top-10\% share \\
\midrule
Arbitrage & Full period & 2,587,619 & 7,453 & 53.60\% & 91.37\% \\
 & 2024Q4 & 225,318 & 914 & 80.32\% & 87.85\% \\
 & 2026Q3 & 380,873 & 3,749 & 50.25\% & 82.79\% \\
\addlinespace[2.5pt]
Sandwich & Full period & 2,558,637 & 670 & 95.84\% & 97.24\% \\
 & 2024Q4 & 468,934 & 207 & 99.52\% & 97.32\% \\
 & 2026Q3 & 126,651 & 133 & 98.14\% & 86.39\% \\
\addlinespace[2.5pt]
Liquidations & Full period & 38,225 & 762 & 76.77\% & 81.83\% \\
 & 2024Q4 & 846 & 98 & 91.84\% & 63.24\% \\
 & 2026Q3 & 1,516 & 134 & 92.08\% & 81.27\% \\
\addlinespace[2.5pt]
All strategies & Deduplicated & 5,041,034 & 8,516 & 68.12\% & 95.33\% \\
\bottomrule
\end{tabular*}
\vspace{3pt}
\begin{minipage}{\linewidth}
\footnotesize Notes: Distinct sender addresses and verified transaction shares across strategies (2024Q4 and 2026Q3 are the first and last complete UTC quarters).
\end{minipage}
\end{table}

Participation moved in opposite directions depending on what each strategy requires beyond code. Distinct active arbitrage addresses roughly quadrupled between the two end quarters, the mean number active on a given day more than tripled, and verified arbitrages per day rose by about two thirds. Distinct active liquidators grew by about a third; the quarterly count moves with liquidation volume, and both fell between the second and third quarters of 2026 as fewer positions crossed their collateral thresholds. Distinct active sandwich addresses, by contrast, fell by more than a third, in the one strategy that requires low-latency order visibility and builder access that cheaper software does not supply.

Execution nonetheless remained concentrated. In arbitrage the share of the fifty most active addresses fell by thirty points between the two end quarters, from about four fifths to about half, yet the top ten percent of addresses still execute more than four fifths of the final quarter\textquotesingle s arbitrages. In sandwich attacks the top fifty\textquotesingle s share stayed close to the whole in both end quarters, and in liquidations the top ten percent execute more than four fifths over the full ledger. Entry thus broadened the base of participants in strategies where code suffices, while the operators who capture execution continue to hold advantages beyond software.

\subsection{Contract deployment and trade execution across addresses}\label{contract-deployment-and-trade-execution-across-addresses}

Across the verified addresses, a few hundred deploy at least one top-level contract over the target period, tens of thousands of creations in all, and more than three fifths of the inspected 2026 runtime bytecodes have distinct code hashes with no lightweight forwarding contracts (minimal proxies) among them, so the deployments are distinct programs. The number of deploying addresses rose in each quarter, more than tripling between the first quarter of 2025 and the third quarter of 2026, while creations per deploying address fell to about a sixth, so total quarterly creations fell by roughly two fifths even as the number of deployers grew. The weekend share of creations fell from about a quarter to about a fifth, below the share a uniform seven-day clock would give (Supplementary Section S1.3).

Deployment and execution are distributed differently across address groups. Ranking the verified addresses by verified transactions over the full ledger, the top fifty execute about two thirds of all transactions and deploy under two percent of the top-level contracts; the remaining addresses, more than 99 percent of the population, execute the rest of the trades and deploy nearly all of the contracts. A leading address deploys about three times as many contracts as a smaller one and is far more likely to deploy at all, about a third of leading addresses against one in sixteen of the rest, while the concentration of executed trades at the top far exceeds the top\textquotesingle s share of creations. Because the strategy ledger that fixes the ranks and the trade shares runs over the full period including the historical support whereas the top-level creation archive for these addresses covers the target period only, each share is computed within its own series. Inside the prospective liquidation cohorts, pooling the outage windows of all in-window episodes, the smaller group deploys about three quarters of outage-window creations and executes under a tenth of outage-window liquidations. Because operators often use one key to deploy a contract and a different hot wallet to call it, and because most of their deployments are non-trading contracts, address-level deployment counts alone cannot isolate trading-software releases. The chain-wide typed creation census, which classifies each contract by its subsequent on-chain role, is needed for that separation. The next section pairs this census with the AIDA outage registry to test that link directly.

\subsection{Deployment around AI outages}\label{deployment-around-ai-outages}

Table 3 and Figure 4 report deployment responses over the daily panel, which contains well over a hundred general API and coding-agent disruptions of 60 minutes to 24 hours spread over about a fifth of panel days, and across matched episode windows, across contract roles and five operator and bot identification definitions.

Executor-release activity falls during the initial days of an AI disruption and rebounds on day three. In the baseline population of 8,997 verified MEV senders and executor deployers, an eight-hour disruption reduces the number of addresses releasing a new executor to 0.67 of its usual rate on day 0 (0.82 on day 1 and 0.66 on day 2) before rebounding to 1.23 on day 3. In the off-chain release pipeline, coding-assistant disruptions delay both releases awaiting final testing on the disruption day and builds that were still in earlier development stages two days before, and the interrupted work then finishes and ships on day three. Summing predicted daily counts over days 0 to 3 leaves about fifteen percent fewer executor-deploying address-days than the zero-exposure path. On isolated episode windows the executor-release ratio falls below one during the outage and rises to about two in the equal-duration recovery window, with the wide intervals that hourly counts of a daily-scale process imply. In the hourly within-day PPML specification across the 15,600-hour target period using the AIDA registry (709 developer-relevant foundation-model disruptions across 17 vendors), private executor creations on Ethereum fall to 0.68 ({[}0.48, 0.94{]}) and distinct addresses creating an executor fall to 0.69 ({[}0.49, 0.97{]}) during OpenAI outage hours, while verified arbitrage execution across all vendor outage hours remains unchanged at 0.99 ({[}0.96, 1.02{]}).

The one-third drop holds across alternative MEV and bot identification methods from the literature. When operators are identified without transaction-level cycle or sandwich rules by the exclusive single-caller bot architecture of \citet{Daian2020FlashBoys}, under the private multi-key bot rule, or in the broad sample of addresses deploying any non-pool, non-protocol contract that executes DEX swaps, an eight-hour disruption reduces deploying addresses on day 0 by between a third and two fifths, with the same day-2 dip and day-3 rebound as the baseline. Each of these day-0 intervals excludes one under both week-cluster and block-bootstrap inference. Restricting strictly to direct MEV sender EOAs without recovering separate deployer keys attenuates the day-0 decline to about a quarter, confirming that attributing factory-created and deployer-key contracts back to their creators sharpens measurement of the development margin.

When we count all persistent contracts deployed by the baseline operators, more than nine tenths of which are non-trading contracts that do not execute MEV or receive swaps, the number of deploying addresses does not fall, with a day-0 ratio near one and a four-day cumulative change whose interval includes no change. Capping creations at 3 per address-day also leaves the ratio near one. The uncapped raw creation counts do dip on disruption days, but this is driven entirely by a handful of addresses that pause bulk non-trading deployment batches in 2026; the per-address cap removes these outliers and the dip disappears. The executor decline is statistically distinct from both benchmarks. Across all DEX trading-contract deployers the day-0 ratio on all contracts is one. For the constructor-executed trades, single-use contracts that execute an identified MEV trade inside their constructor, activity during disruptions moves in the opposite direction from executor development, running slightly above its usual rate for active senders on day 0. Even when a bot uses the contract-creation opcode as its execution vehicle, run-time trading continues through AI outages while the release of persistent trading executors pauses.

We also check concurrent cloud incidents, late-sample follow-up length and pre-onset timing. First, AI-provider incidents sometimes coincide with disruptions at public cloud providers that could slow deployment directly; controlling for concurrent cloud-provider disruption hours on the same day leaves the executor decline unchanged. Second, a contract is classified as a trading executor from its subsequent swaps, so contracts deployed near the end of the sample have a shorter window in which to qualify, and a decline concentrated in late 2026 could reflect censoring instead of behavior. Comparing the common January-to-October months of 2025 and 2026, contracts from the 2025 window have more than eleven months of follow-up, and the day-0 declines in the two years are statistically indistinguishable, so shorter late-sample follow-up does not drive the estimate. Third, in exact UTC windows before onset, executor-deploying addresses show no consistent movement three to six hours, one to three hours or one hour before onset, each on small hourly counts with an interval that includes one (Figure 4, Panel B).

\begin{table}[htbp]
\centering
\caption{Deployment around AI disruptions across identification methods and contract roles.}
\label{tab_population_deployment}
\footnotesize
\renewcommand{\arraystretch}{1.10}
\setlength{\tabcolsep}{2.5pt}
\begin{tabular*}{\linewidth}{@{\extracolsep{\fill}}lrrrr@{}}
\toprule
Specification & Estimate & 95\% CI & Block-bootstrap 95\% CI & p \\
\midrule
\multicolumn{5}{@{}l@{}}{\textbf{Panel A: Baseline MEV population (8,997 operators), daily PPML (8h outage)}} \\[1pt]
Day 0 (disruption day) & 0.67 & {[}0.49, 0.92{]} & {[}0.45, 0.91{]} & 0.012 \\
Day 1 & 0.82 & {[}0.57, 1.17{]} & n.a. & 0.276 \\
Day 2 & 0.66 & {[}0.51, 0.86{]} & n.a. & 0.002 \\
Day 3 & 1.23 & {[}0.90, 1.69{]} & n.a. & 0.197 \\
Cumulative days 0-3 (\%) & -15.3 & {[}-30.6, -0.1{]} & {[}-27.0, 3.5{]} & 0.049 \\
Concurrent cloud-disruption control (day 0) & 0.66 & {[}0.46, 0.94{]} & n.a. & 0.023 \\
2026 vs. 2025 follow-up contrast (day 0) & 0.95 & {[}0.53, 1.69{]} & n.a. & 0.850 \\
\addlinespace[3pt]
\multicolumn{5}{@{}l@{}}{\textbf{Panel B: Alternative operator definitions, executor releases on day 0}} \\[1pt]
Single-caller bots (2,387 operators) & 0.62 & {[}0.44, 0.88{]} & {[}0.37, 0.88{]} & 0.007 \\
Private bots <=3 callers (3,054 operators) & 0.65 & {[}0.44, 0.97{]} & {[}0.39, 0.94{]} & 0.035 \\
All DEX trading deployers (18,970 operators) & 0.64 & {[}0.46, 0.89{]} & {[}0.42, 0.89{]} & 0.007 \\
Direct MEV sender EOAs (8,516 operators) & 0.73 & {[}0.53, 1.01{]} & {[}0.46, 0.99{]} & 0.059 \\
\addlinespace[3pt]
\multicolumn{5}{@{}l@{}}{\textbf{Panel C: Contract-role specificity benchmarks and isolated episode windows}} \\[1pt]
All persistent contracts, days 0-3 (\%) & -2.0 & {[}-7.4, 3.5{]} & {[}-6.7, 2.5{]} & 0.485 \\
All persistent contracts, day 0 & 0.94 & {[}0.84, 1.05{]} & {[}0.85, 1.05{]} & 0.248 \\
Creations capped at 3/day, day 0 & 0.97 & {[}0.89, 1.06{]} & {[}0.90, 1.07{]} & 0.578 \\
Uncapped creations, day 0 & 0.78 & {[}0.62, 0.98{]} & {[}0.40, 1.02{]} & 0.035 \\
Contrast: executors vs. persistent contracts (day 0) & 0.72 & {[}0.51, 1.00{]} & n.a. & 0.048 \\
Runtime constructor-executed trades, day 0 & 1.11 & {[}1.00, 1.23{]} & {[}0.98, 1.24{]} & 0.050 \\
Isolated windows: executor releases (during / recovery) & 0.83 / 2.03 & {[}0.53, 1.30{]} / {[}0.51, 5.07{]} & n.a. & n.a. \\
Isolated windows: persistent contracts (during / recovery) & 0.96 / 1.17 & {[}0.85, 1.10{]} / {[}1.00, 1.37{]} & n.a. & n.a. \\
\addlinespace[3pt]
\multicolumn{5}{@{}l@{}}{\textbf{Panel D: Hourly within-day AIDA panel and Base L2 cross-chain panel}} \\[1pt]
Ethereum L1: private executors (OpenAI outage hours) & 0.68 & {[}0.48, 0.94{]} & n.a. & 0.020 \\
Base L2: operator top-level creations (OpenAI outage hours) & 0.76 & {[}0.59, 0.99{]} & n.a. & 0.042 \\
Base L2: distinct operators deploying executor (day 2) & 0.88 & {[}0.78, 0.98{]} & n.a. & n.a. \\
\bottomrule
\end{tabular*}
\vspace{3pt}
\begin{minipage}{\linewidth}
\footnotesize Notes: Daily PPML over 650 days (62 exposed week clusters; 95\% CIs clustered by Monday-anchored week) for an 8h Anthropic or OpenAI disruption; isolated-window rows report weekly block-bootstrap 95\% CIs across 159 outage and 88 recovery episodes; hourly PPML rows control for UTC-date and hour-of-day fixed effects across 15,600 hours with UTC-date-clustered 95\% CIs.
\end{minipage}
\end{table}

\begin{figure}
\centering
\pandocbounded{\includegraphics[keepaspectratio,alt={Deployment response to AI service disruptions for the chain-wide MEV population. Panel A plots daily PPML rate ratios for an eight-hour disruption across days 0 to 3 with 95 percent week-cluster intervals for addresses releasing a trading executor, addresses releasing any persistent contract, uncapped creations, and creations capped at three per day. Panel B plots isolated episode-window rate ratios with weekly block-bootstrap intervals from six hours before onset through the equal-duration recovery window.}]{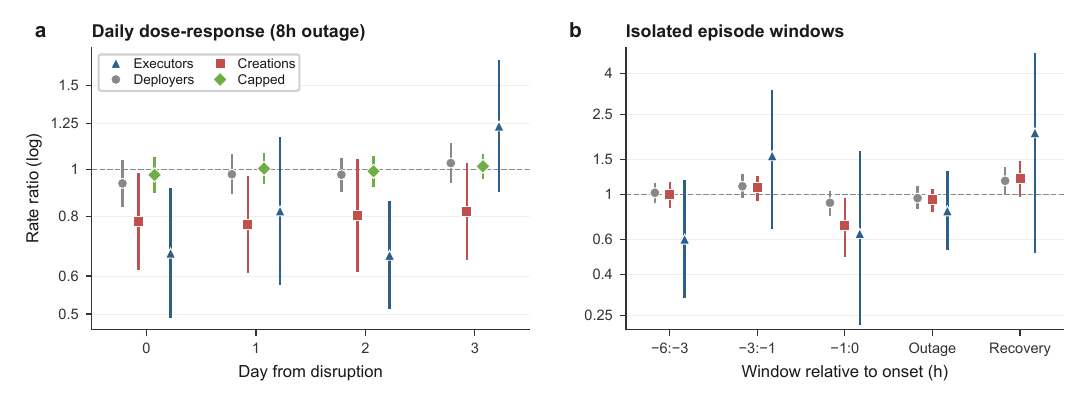}}
\caption{Deployment response to AI service disruptions for the chain-wide MEV population. Panel A plots daily PPML rate ratios for an eight-hour disruption across days 0 to 3 with 95 percent week-cluster intervals for addresses releasing a trading executor, addresses releasing any persistent contract, uncapped creations, and creations capped at three per day. Panel B plots isolated episode-window rate ratios with weekly block-bootstrap intervals from six hours before onset through the equal-duration recovery window.}
\end{figure}

Code updates via EIP-7702 account delegation, which attaches or swaps executable code on an existing wallet address (over a thousand executors), and implementation swaps behind proxy contracts do not emit contract-creation transactions and remain outside the creation census; the factory census covers the factories through which private executors were created, not all factories on Ethereum; and an on-chain address represents a cryptographic key, not a consolidated firm. Sparse creation counts within the prospective liquidation cohorts are reported in Supplementary Section S7.

\subsection{Execution around AI outages}\label{execution-around-ai-outages}

\subsubsection{Arbitrage and sandwich execution on matched episodes}\label{arbitrage-and-sandwich-execution-on-matched-episodes}

The deployment results measure new additions to the stock of trading software; we next test whether the bots already operating continue executing during the disruption. Atomic arbitrage and sandwich bots execute hundreds of transactions per hour across market states, so counting all verified arbitrage and sandwich transactions inside each incident window and its references gives a high-frequency test of whether aggregate market execution slows when a provider goes down. On the nearby-day episodes in the baseline cloud-screened sample, arbitrage and sandwich attacks both run within a few percent of their reference rates, while the extended full-period sample and the same-weekday calendar give ratios at or above these (Supplementary Section S2.2). Seven of the eight outage-window confidence intervals include one; the exception, same-weekday arbitrage in the baseline sample, lies above one, and none lies below one. The nearby-day intervals exclude declines larger than about a tenth for arbitrage and smaller still for sandwich attacks.

\subsubsection{Liquidations}\label{liquidations}

Liquidations differ structurally from DEX swaps because a borrower\textquotesingle s position must cross a protocol collateral threshold before a liquidation can occur, so window counts reflect discrete price-triggered opportunities as well as bot execution. Table 4 reports the liquidation estimates and the smaller group\textquotesingle s share changes in both the baseline cloud-screened sample (through late April 2026; Section 3.2) and the extended full-period sample through October 2026, across both reference calendars. Figure 5 places the liquidation transaction ratios beside the top-level contract creations and the creating senders of the same cohorts in both samples, so that the effect of the cloud screen can be read across all three outcomes.

\begin{table}[htbp]
\centering
\caption{Liquidation trading activity and smaller-group share changes across samples and calendars.}
\label{tab_liquidation_primary}
\scriptsize
\renewcommand{\arraystretch}{1.10}
\setlength{\tabcolsep}{2.5pt}
\begin{tabular*}{\linewidth}{@{\extracolsep{\fill}}llrrrr@{}}
\toprule
Sample & Reference calendar & Observed / reference & Estimate & 95\% interval & n \\
\midrule
\multicolumn{6}{@{}l@{}}{\textbf{Panel A: Verified liquidation transactions (rate ratio)}} \\[1pt]
Extended & Nearby days, all & 697 / 1,366.05 & 0.51 & {[}0.18, 1.51{]} & 198 \\
 & Same weekday, all & 183 / 768.67 & 0.24 & {[}0.11, 0.65{]} & 98 \\
 & Nearby days, balanced & 123 / 374.18 & 0.33 & {[}0.13, 1.00{]} & 73 \\
 & Same weekday, balanced & 123 / 334.83 & 0.37 & {[}0.22, 0.71{]} & 73 \\
\addlinespace[2.5pt]
Baseline & Nearby days, all & 329 / 1,303.03 & 0.25 & {[}0.11, 0.61{]} & 156 \\
 & Same weekday, all & 165 / 746.67 & 0.22 & {[}0.10, 0.62{]} & 84 \\
 & Nearby days, balanced & 106 / 358.40 & 0.30 & {[}0.11, 0.97{]} & 64 \\
 & Same weekday, balanced & 106 / 327.33 & 0.32 & {[}0.18, 0.61{]} & 64 \\
\addlinespace[3pt]
\multicolumn{6}{@{}l@{}}{\textbf{Panel B: Smaller-group share change (percentage points)}} \\[1pt]
Extended & Nearby days, all & defined in 87 & 2.87 & {[}-5.95, 12.44{]} & 87 \\
 & Same weekday, all & defined in 46 & 3.23 & {[}-6.35, 14.71{]} & 46 \\
\addlinespace[2.5pt]
Baseline & Nearby days, all & defined in 69 & 4.22 & {[}-5.10, 14.32{]} & 69 \\
 & Same weekday, all & defined in 42 & -2.46 & {[}-8.81, 4.90{]} & 42 \\
 & Nearby days, balanced & defined in 30 & -4.41 & {[}-17.43, 7.55{]} & 30 \\
 & Same weekday, balanced & defined in 30 & -1.18 & {[}-10.82, 9.18{]} & 30 \\
\bottomrule
\end{tabular*}
\vspace{3pt}
\begin{minipage}{\linewidth}
\footnotesize Notes: Prospective liquidation cohorts (ranks 1-50 leading; 51+ smaller); 4,000 Monday-week bootstrap draws.
\end{minipage}
\end{table}

Matched liquidation counts during outages are lower than their references across all specifications. In the baseline cloud-screened sample the nearby-day and same-weekday ratios both sit near a quarter of the reference rate, and the extended-sample and balanced-sample estimates are in Table 4. On the nearby-day calendar the decomposition \(R_Q = R_A R_I\) splits the baseline shortfall into participation and intensity, as fewer liquidators were active and each executed fewer liquidations, with participation the larger component, and the other calendar and sample give similar splits. The extended sample shows a milder shortfall of about half, and the difference between the two samples is a composition effect of the post-April-2026 episodes excluded by the cloud screen, since the retained episodes keep identical counts and references. The nearby-day interval in the baseline sample implies a decline of at least about two fifths.

\begin{figure}
\centering
\pandocbounded{\includegraphics[keepaspectratio,alt={Three-panel comparison of liquidation trading activity, top-level contract creations and creating-sender window counts in the extended full-period and baseline cloud-screened samples. Each panel reports rate ratios and weekly 95 percent bootstrap intervals across nearby-day and same-weekday reference calendars, across all matched episodes and in the balanced sample matched on both calendars; Table 4 reports the episode counts.}]{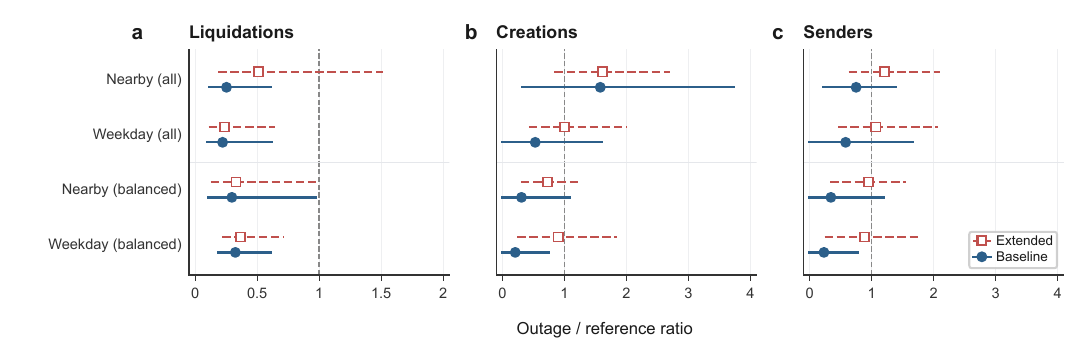}}
\caption{Three-panel comparison of liquidation trading activity, top-level contract creations and creating-sender window counts in the extended full-period and baseline cloud-screened samples. Each panel reports rate ratios and weekly 95 percent bootstrap intervals across nearby-day and same-weekday reference calendars, across all matched episodes and in the balanced sample matched on both calendars; Table 4 reports the episode counts.}
\end{figure}

\subsubsection{Competitive shares}\label{competitive-shares}

The smaller group\textquotesingle s share of liquidations does not detectably shift toward leading addresses in any specification (Table 4). In the baseline cloud-screened sample the duration-weighted share change is a few percentage points in the smaller group\textquotesingle s favor relative to nearby days and a few points against it relative to same weekdays, and the weekly and fortnight intervals across all matched episodes, in the balanced sample and in both the baseline and extended samples all include zero. Across all matched episodes the intervals exclude losses in the smaller group\textquotesingle s share larger than five to nine percentage points on either calendar, with wider intervals in the 30-episode balanced subsample, among cohort addresses on episodes with positive counts on both sides.

\subsubsection{Market context}\label{market-context}

Because liquidation volume depends on collateral-price moves, Table 5 partitions all verified Aave and Compound liquidations into cohort and outside-cohort activity over the same windows.

\begin{table}[htbp]
\centering
\caption{Decoded lending-market liquidation context and outside-cohort partition.}
\label{tab_market_context}
\scriptsize
\renewcommand{\arraystretch}{1.10}
\setlength{\tabcolsep}{2.5pt}
\begin{tabular*}{\linewidth}{@{\extracolsep{\fill}}lllrrrr@{}}
\toprule
Sample & Calendar & Support (n) & All-market obs / ref & All-market ratio & Cohort ratio & Outside-cohort ratio \\
\midrule
Extended & Nearby days & All (198) & 834 / 1,651.48 & 0.51 & 0.51 & 0.48 \\
 &  & Balanced (73) & 174 / 441.53 & 0.39 & 0.33 & 0.76 \\
 & Same weekday & All (98) & 249 / 891.17 & 0.28 & 0.24 & 0.54 \\
 &  & Balanced (73) & 174 / 380.17 & 0.46 & 0.37 & 1.12 \\
\addlinespace[2.5pt]
Baseline & Nearby days & All (156) & 431 / 1,559.08 & 0.28 & 0.25 & 0.40 \\
 &  & Balanced (64) & 152 / 414.35 & 0.37 & 0.30 & 0.82 \\
 & Same weekday & All (84) & 225 / 867.17 & 0.26 & 0.22 & 0.50 \\
 &  & Balanced (64) & 152 / 371.17 & 0.41 & 0.32 & 1.05 \\
\bottomrule
\end{tabular*}
\vspace{3pt}
\begin{minipage}{\linewidth}
\footnotesize Notes: Partition of all 38,225 verified Aave V2, Aave V3 and Compound V2 liquidations into prospective cohort and outside-cohort activity.
\end{minipage}
\end{table}

The all-market aggregate lies below one across all matched episodes in both samples, and because the cohort accounts for between seventy and ninety percent of decoded liquidations in incident and reference windows, the aggregate largely restates the cohort result. In the balanced sample the outside-cohort remainder sits below one relative to nearby days and at or slightly above one relative to same weekdays, in the baseline and extended samples alike (Table 5), while the cohort\textquotesingle s own balanced-sample ratios differ little between calendars. Thus the activity of non-cohort liquidators in the same windows is sensitive to which days serve as the reference, whereas a genuine outage effect on cohort bots would appear consistently regardless of the reference calendar.

\subsubsection{Timing profiles, placement placebo and recorded strata}\label{timing-profiles-placement-placebo-and-recorded-strata}

Figure 6 and Supplementary Table S2 report thirty diagnostic cases in the baseline cloud-screened sample across five event windows on two calendars (across all episodes matched for each window and in the balanced sample matched across all five windows), a shared-reference placement placebo, and recorded episode strata.

\begin{figure}
\centering
\pandocbounded{\includegraphics[keepaspectratio,alt={Liquidation timing profiles in the baseline cloud-screened sample across five event windows under nearby-day and same-weekday reference calendars. Panels A and B plot transaction and active-operator rate ratios across all episodes matched for each window; Panels C and D plot the corresponding profiles restricted to the balanced sample of episodes matched across all five windows, with episode counts in Supplementary Table S2.}]{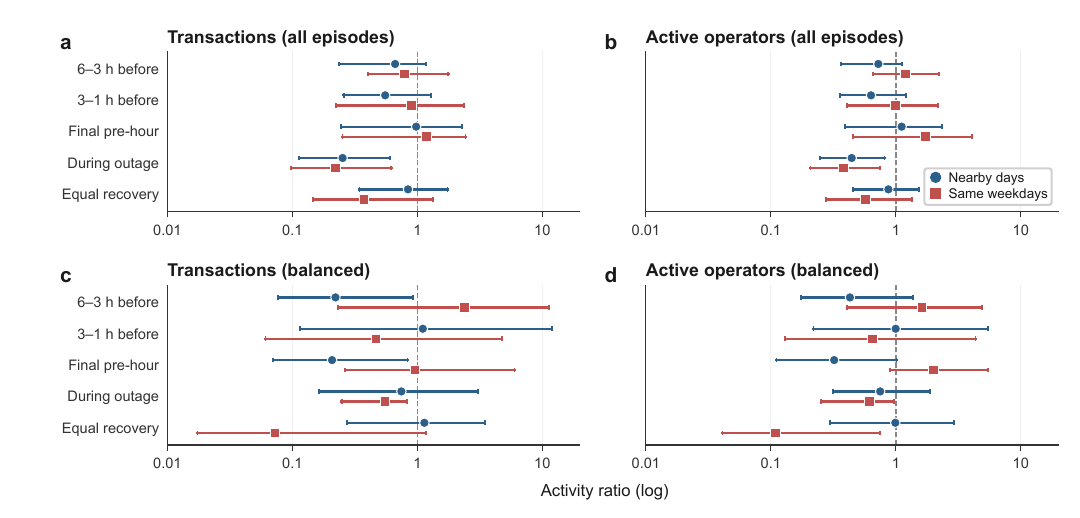}}
\caption{Liquidation timing profiles in the baseline cloud-screened sample across five event windows under nearby-day and same-weekday reference calendars. Panels A and B plot transaction and active-operator rate ratios across all episodes matched for each window; Panels C and D plot the corresponding profiles restricted to the balanced sample of episodes matched across all five windows, with episode counts in Supplementary Table S2.}
\end{figure}

Across all matched episodes (Figure 6, Panels A and B) the three pre-onset windows have intervals that include one on both calendars, the outage window reproduces the primary ratios, and the equal-duration recovery window lies below one on both calendars, mildly so relative to nearby days and sharply relative to same weekdays. Because not all episodes have clean references in every window, the profile across all matched episodes draws on a different set of episodes in each window; Panels C and D therefore hold the episode set fixed by restricting to episodes matched across all five windows. In the nearby-day balanced sample, two of the three pre-onset windows lie below one while the outage window and the recovery window include one, showing no contraction that begins at onset and ends at recovery; in the same-weekday balanced sample, the outage window is the only one of the five whose interval excludes one.

The placement placebo compares the outage window and the pre-outage day placebo, the clean same-clock window one day before onset, against the same references drawn from two to five days earlier. On the balanced-sample episodes the outage-window interval excludes one and the placebo-window interval does not, and the paired contrast between the two arms indicates that the outage window holds about half as many liquidations as the window one day earlier, with an interval that includes equality under weekly and fortnight clusters alike (Supplementary Table S2). Episodes with no overlapping provider or infrastructure disruption and episodes confined to API and coding-assistant surfaces both give point estimates well below one on both calendars, with the same-weekday intervals excluding one. Together, the diagnostics rule out an onset-to-recovery contraction on the nearby-day calendar and show that the outside-cohort remainder depends on the reference calendar, while the paired placebo leaves open a shortfall on the outage day itself.

\subsection{Cross-chain extensions on Base and Solana and quota announcements}\label{cross-chain-extensions-on-base-and-solana-and-quota-announcements}

On the 22-month Base panel (Section 3.4), the separation between off-chain development and on-chain execution reproduces the Ethereum findings (Table 3 and Supplementary Section S3.2). Because low Layer-2 gas costs allow deployed contracts to spawn over a million internal factory clones during routine operation, separating operator-signed top-level creations and per-operator counts from uncapped internal factory spawns isolates the development margin. In the hourly within-day PPML model under AIDA, top-level contract creations signed directly by Base MEV operators (93,694 creations by 2,888 deploying operators) fall to 0.76 ({[}0.59, 0.99{]}) during OpenAI outage hours, and top-level creations in the searcher creation-trace archive fall to 0.73 ({[}0.56, 0.96{]}) during GitHub Copilot disruptions. In the daily panel on Base, distinct operators deploying an executor fall to 0.88 ({[}0.78, 0.98{]}) two days after an eight-hour Anthropic or OpenAI disruption, daily sums of executor creations capped at three per operator-hour fall to 0.89 ({[}0.78, 1.00{]}), and first-time operator caller addresses invoking private executors fall to 0.88 ({[}0.78, 0.99{]}) at the same two-day lag, matching the two-day pipeline delay observed on Ethereum. Meanwhile, across all vendor outage hours, hourly executor creations remain near one both when capped at three per operator-hour (1.00, {[}0.92, 1.09{]}) and when uncapped internal factory spawns are included (1.04, {[}0.94, 1.15{]}), and live trade execution across 26,165,900 Base arbitrage executor transactions (1.00, {[}0.97, 1.02{]}) and all 1,099,593,191 Base pool swaps (1.00, {[}0.98, 1.01{]}) continues without interruption.

The Solana extension establishes the mechanics and the archive cost of a non-EVM cross-chain panel. We verify loader-level program deployment and upgrade decoding, fix the first 90-day discovery interval at an exact cutoff slot, and verify about a thousand Raydium concentrated-liquidity swaps across four historical strata, about half of them routed through the Jupiter aggregator. Because swap discovery on Solana requires tens of kilobytes of archive reads per slot, a single 90-day cohort window would need close to a terabyte of archive reads, so we report the bounded evidence without extrapolating a Solana outage panel (Supplementary Section S3.1). Attribution also differs in kind because a shared router program is not an individual trader\textquotesingle s software, so program upgrades cannot stand in for Ethereum-style deployments without resolving the caller.

Figure 7 and Supplementary Table S4 report daily deployments by six frozen liquidation cohorts around six public AI quota and access announcements in 2026. Creation counts are sporadic and driven by occasional multi-contract days of individual addresses; after cloud screening only a handful of display dates, all around the 2 February OpenAI policy expansion, have at least two clean same-weekday baselines, and on those dates observed creations sit at or below the baseline mean. The profiles provide no support for a post-announcement deployment burst.

\begin{figure}
\centering
\pandocbounded{\includegraphics[keepaspectratio,alt={Daily top-level contract creations for six prospective liquidation cohorts around public 2026 AI access and quota announcements. Each panel spans 7 days before through 14 days after the announcement, marking the announcement date and same-weekday baseline means where clean controls exist.}]{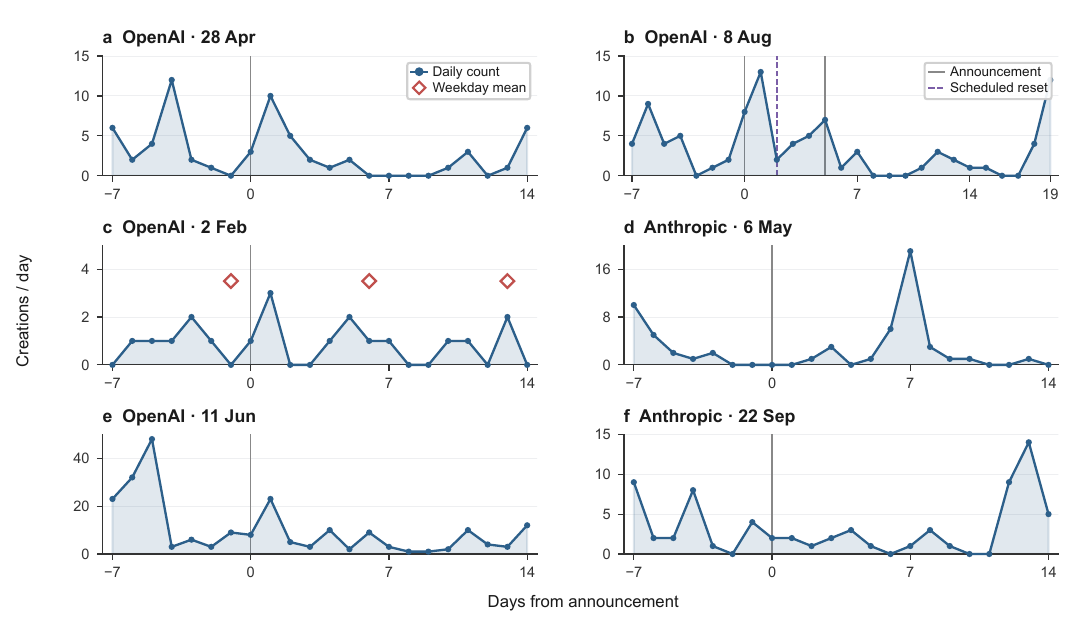}}
\caption{Daily top-level contract creations for six prospective liquidation cohorts around public 2026 AI access and quota announcements. Each panel spans 7 days before through 14 days after the announcement, marking the announcement date and same-weekday baseline means where clean controls exist.}
\end{figure}

\FloatBarrier

\section{Discussion}\label{discussion}

\subsection{Entry, concentration and the role of cheaper software}\label{entry-concentration-and-the-role-of-cheaper-software}

The entry and concentration patterns correspond to what each strategy requires beyond code. Atomic arbitrage and liquidations identify candidate opportunities from public pool state and borrower health factors, although winning execution still depends on financing, latency and transaction placement. Sandwich attacks require low-latency sight of pending user orders and placement of the attacker\textquotesingle s bundle with a block builder, inputs that coding tools do not supply. Entry expanded in the two public-state strategies and contracted in the one that requires private order flow, while deployments per deploying address fell and shifted toward working-week hours (Table 2 and Section 4.3). The combination of broader entry in public-state strategies, fewer deployments per deploying address, and weekday concentration is what lower software-production costs predict.

That entry expanded while execution stayed concentrated sharpens the interpretation. The leading addresses continue to capture the large majority of verified trades, even as the deployer base expanded severalfold (Sections 4.2 and 4.3). Cheaper software thus lowers the barrier to building and releasing MEV contracts, while private order flow, builder access, latency and capital continue to determine who captures execution.

Two alternative explanations deserve consideration. If operators rotated addresses more frequently over time, distinct-address counts would rise without an expansion in participants; the daily-active measure, which counts an address only on days it trades, rises by a similar factor and limits that concern. Changes in the opportunity set, such as the growth of Uniswap V4 pools and new lending markets, would draw entrants regardless of software costs; we observe entry in both strategies that read public state and exit in the one that requires private order flow, which an expansion of trading venues alone does not predict. Because the chain does not record which operators use which coding tools, the cross-strategy entry pattern is indirect evidence; the outage design provides the direct test.

\subsection{Off-chain development and on-chain execution}\label{off-chain-development-and-on-chain-execution}

The AIDA registry allows a direct test of whether the addresses that run these strategies keep releasing code and executing trades when a named AI provider becomes unavailable. Verified arbitrage and sandwich execution on the matched episodes show no detectable decline (Section 4.5.1), and the five-window profile and the outside-cohort partition show that the lower liquidation counts do not track outage onset and offset in the way a run-time dependence would, where one would expect a drop beginning at onset and recovery after resolution. Because Ethereum blocks arrive at twelve-second intervals and builders close their bundle auctions a fraction of a second before the slot deadline, a bot that wins execution decides and submits in milliseconds against current state, whereas a cloud language-model call takes several seconds. Moreover, constructor-executed trades, the one form of execution that leaves a creation record, continue at or slightly above their usual rate on disruption days, and the smaller group\textquotesingle s share of liquidations is stable, so neither execution volume nor its allocation moved at the scale that large cross-operator substitution would produce.

Observing deployment and execution for the same operators addresses a central identification challenge of outage studies: distinguishing whether the affected population actually uses the disrupted service. If we observed only that arbitrage and sandwich execution continue during AI disruptions, one could object that MEV searchers do not use commercial AI coding services. On the deployment margin, however, executor releases decline on the disruption day across all five identification methods and follow the multi-day dip-and-rebound of a staged software pipeline, whereas non-trading creations by the same operators remain flat and concurrent cloud disruptions or late-sample censoring do not account for the drop (Section 4.4 and Table 3). The same separation holds on Base once operator-signed creations and executor-deploying operators are separated from routine internal factory clones (Section 4.6). Across both chains, MEV developers rely on AI coding assistants when writing, testing and shipping new trading contracts, while their already-deployed bots execute on-chain without run-time dependence on the disrupted AI services, matching the offline synthesis-and-fork-test workflow of \citet{Wen2026EVAGE} and the finding by \citet{Brynjolfsson2025Work} that customer-support agents retain their productivity gains during assistant outages once AI-aided routines have been learned. The stability of smaller operators\textquotesingle{} competitive share during outages confirms that outages do not tilt live capture toward larger incumbents (Table 4).

\subsection{The liquidation shortfall}\label{the-liquidation-shortfall}

Liquidation counts during outages lie below their reference rates across the primary specifications (Section 4.5.2 and Table 4), and the diagnostics help distinguish among timing, opportunity arrivals and cohort composition as candidate explanations. Liquidations arrive in bursts when collateral prices move, and in the balanced sample the outside-cohort partition shows fewer liquidations by non-cohort addresses relative to nearby days than relative to same weekdays (Table 5), so outside-cohort activity in those windows depends on the reference calendar in a way a bot-level outage effect should not. On the balanced five-window profile under the nearby-day calendar, two pre-onset windows already lie below one while the outage and recovery windows include one (Figure 6), ruling out a contraction that begins at onset and ends at recovery on that calendar, even though the outage window remains below one on the same-weekday profile. Conversely, in the shared-reference placement placebo the outage window records fewer liquidations than the clean window one day earlier, a paired contrast whose interval includes equality (Supplementary Table S2) yet which a pure reference-timing explanation does not predict. The candidate explanations that survive are lumpiness of liquidation opportunities coinciding with outage placement, residual overlap with infrastructure incidents that the public cloud-status records do not capture, and aging of cohorts between their discovery windows and the outage. Resolving among them requires an independent measure of opportunities, for instance the on-chain oracle price path that triggers liquidations, and placebo outages constructed with the cohort rules; we report the shortfall as unresolved.

\subsection{Measurement scope and limitations}\label{measurement-scope-and-limitations}

Transaction-level identification separates strategy execution from unrelated order flow. Classifying contracts by pool-count medians or block positions would have counted routing and victim flow as extraction; the sequential-cycle, sandwich-band and decoded-liquidation rules link each identified transaction to its decoded swaps or liquidation events. Reference-calendar choice and the coverage of the cloud screen change estimates through episode composition, which is why we report both calendars in the balanced sample and both the baseline and extended samples.

Several measurement boundaries apply to the interpretation. Addresses are transaction senders, and one organization may operate several; the deployment-versus-execution contrast is a contrast between address groups, and a dedicated deployer of a large operator could sit in the smaller group. Creations measure deployment of new contracts, and the role of a created contract is read from what it later does, which separates executors from unrelated code while leaving in-place wallet code delegation under EIP-7702, upgrades behind proxies, and creations through factories other than private-executor factories outside the creation census. Finally, the nine-signature pool-swap filter bounds recall at the source (Section 3.1 and Supplementary Section S1.2), so our arbitrage counts are lower bounds and any cycle routed partly through an unlisted pool or non-pool venue is counted as broken.

\FloatBarrier

\section{Conclusion}\label{conclusion}

Generative AI has made trading software cheaper to write, but no systematic dataset has recorded when these services fail. We introduced AIDA, a multi-vendor registry of 5,868 AI outage incidents across 30 providers and 17 foundation-model vendors, and paired it with transaction-level MEV identification over the full Ethereum swap and liquidation ledger (5,041,034 strategy transactions by 8,516 addresses) and a 22-month Base Layer-2 panel (1,099,593,191 swaps and 29,120 private executors).

Two findings emerge. First, entry into MEV expanded where code and public state suffice. Active arbitrage addresses quadrupled between the fourth quarter of 2024 and the third quarter of 2026, liquidators grew by a third, and the number of addresses deploying contracts in a quarter more than tripled, with deployments concentrating on weekdays; sandwich attackers, who need private order flow and builder access, declined. Execution nonetheless remained concentrated, as the fifty most active addresses execute about two thirds of verified trades, the expanding pool of deployers releases fewer contracts per address, and the top decile of arbitrage addresses still executes more than four fifths of late-2026 arbitrages.

Second, AI outages stall new trading code while deployed bots keep trading across both chains. On Ethereum, the number of addresses releasing a private executor contract falls to 0.67 of its usual rate on the day of an eight-hour disruption (0.69 for deploying addresses and 0.68 for executor creations during OpenAI outage hours) and 0.66 two days later before rebounding on day three, a decline that is statistically distinct from the flat response of all persistent contract creations by the same operators and holds across five identification methods, both subsample years, and a direct cloud-disruption control. On Base, operator-signed top-level creations fall to 0.76 ({[}0.59, 0.99{]}) during OpenAI outage hours and executor-deploying operators fall to 0.88 ({[}0.78, 0.98{]}) two days after an eight-hour Anthropic or OpenAI disruption. Across both chains, live arbitrage and sandwich execution and non-trading or automated factory creations remain unchanged, as do constructor-executed trades and smaller operators\textquotesingle{} liquidation shares on Ethereum.

Cheaper code thus lowers the barrier to building MEV contracts, while private order flow, builder access and execution speed continue to determine who captures opportunities on-chain. AIDA and the paired on-chain identification framework provide a reusable infrastructure for studying how AI service disruptions propagate through software-intensive markets.

\section*{Declarations}\label{declarations}
\addcontentsline{toc}{section}{Declarations}

Data and code availability: The AI Incident and Disruption Archive (AIDA) dataset is hosted on Hugging Face at \url{https://huggingface.co/datasets/venvoo/aida}, and the analysis code and replication package are available on GitHub at \url{https://github.com/dthinkr/agentic-mev}. The replication package regenerates the reported estimates, diagnostics and figures from the transaction-level datasets without network access. Source reconstruction uses Ethereum and Base records from Envio HyperSync and the SQD Network archive, Solana records from the SQD mainnet archive, and the providers\textquotesingle{} public status histories.

Funding: none.

Competing interests: none.

Use of generative AI: The author used Claude Code, with GPT and Claude models, for drafting and editing text, developing and checking analysis code, and preparing data-visualisation code. The author takes responsibility for the analysis and manuscript.

\bibliographystyle{elsarticle-harv}
\bibliography{references}

\clearpage
\appendix
\subsection*{S1. Ledger coverage, collector audit and quarterly series}\label{s1.-ledger-coverage-collector-audit-and-quarterly-series}
\addcontentsline{toc}{subsection}{S1. Ledger coverage, collector audit and quarterly series}

\subsubsection*{S1.1 Ethereum swap classification by block interval}\label{s1.1-ethereum-swap-classification-by-block-interval}
\addcontentsline{toc}{subsubsection}{S1.1 Ethereum swap classification by block interval}

The Ethereum swap ledger tiles blocks {[}20,680,663, 26,179,842) across 14 contiguous intervals without gaps, covering the 118-day historical support period from 5 September 2024 and the 22-month target period from 1 January 2025 through 12 October 2026. Table S1a reports observed transactions, verified atomic arbitrages, verified sandwich transactions (and patterns) and unresolved dense-router transactions in each interval.

\par\medskip\noindent\begin{minipage}{\linewidth}
\textbf{Table S1a.} Ethereum swap ledger coverage and identified MEV transactions by block interval.\par\smallskip
\scriptsize
\renewcommand{\arraystretch}{1.10}
\setlength{\tabcolsep}{2.5pt}
\begin{tabular*}{\linewidth}{@{\extracolsep{\fill}}lrrrrr@{}}
\toprule
Block interval & Blocks & Observed txs & Arbitrage txs & Sandwich txs (patterns) & Unresolved txs \\
\midrule
{[}20680663, 21525891) & 845,228 & 27,948,257 & 339,682 & 670,361 (495,646) & 0 \\
{[}21525891, 21872030) & 346,139 & 9,701,248 & 88,124 & 182,781 (109,504) & 0 \\
{[}21872030, 22606143) & 734,113 & 20,385,231 & 262,247 & 382,152 (223,496) & 1 \\
{[}22606143, 23264566) & 658,423 & 22,187,023 & 246,835 & 391,546 (238,184) & 0 \\
{[}23264566, 23914921) & 650,355 & 18,619,766 & 312,164 & 231,270 (127,737) & 0 \\
{[}23914921, 24587538) & 672,617 & 19,329,858 & 404,427 & 229,094 (125,332) & 3 \\
{[}24587538, 24809732) & 222,194 & 5,927,003 & 109,155 & 68,019 (36,687) & 0 \\
{[}24809732, 25025078) & 215,346 & 7,837,110 & 133,183 & 131,150 (83,181) & 0 \\
{[}25025078, 25247462) & 222,384 & 7,500,681 & 154,802 & 84,565 (47,550) & 0 \\
{[}25247462, 25462599) & 215,137 & 6,068,913 & 124,878 & 55,005 (28,805) & 0 \\
{[}25462599, 25692173) & 229,574 & 5,834,818 & 124,211 & 44,898 (23,599) & 21 \\
{[}25692173, 25907400) & 215,227 & 5,876,246 & 134,475 & 42,360 (22,116) & 17 \\
{[}25907400, 26179609) & 272,209 & 8,105,414 & 153,184 & 45,392 (23,567) & 0 \\
{[}26179609, 26179842) & 233 & 8,399 & 252 & 44 (22) & 0 \\
\textbf{Total {[}20680663, 26179842)} & \textbf{5,499,179} & \textbf{165,329,967} & \textbf{2,587,619} & \textbf{2,558,637} & \textbf{42} \\
\bottomrule
\end{tabular*}
\end{minipage}\par\medskip

\subsubsection*{S1.2 Coverage of the pool-swap protocol filter}\label{s1.2-coverage-of-the-pool-swap-protocol-filter}
\addcontentsline{toc}{subsubsection}{S1.2 Coverage of the pool-swap protocol filter}

The swap collector retains all logs of any transaction that emits at least one of nine pool-level swap event signatures (Uniswap V2, V3 and V4, PancakeSwap V3, Algebra, Balancer V2 and two Curve exchange events). Auditing one unfiltered day of Ethereum logs (15 June 2026) against all pool-level swap events outside that list shows that missed pool-swap transactions represent 1.4 percent of the selected count and transactions carrying an unread leg on an unlisted pool represent 4.1 percent (Table S1b). Non-pool order-flow events (request-for-quote fills, intent settlements and bridges) fall outside the public-pool arbitrage definition.

\par\medskip\noindent\begin{minipage}{\linewidth}
\textbf{Table S1b.} Pool-level swap events outside the baseline protocol filter on the audited day.\par\smallskip
\footnotesize
\renewcommand{\arraystretch}{1.10}
\setlength{\tabcolsep}{2.5pt}
\begin{tabular*}{\linewidth}{@{\extracolsep{\fill}}lrrrr@{}}
\toprule
Pool-level swap event outside baseline list & Logs & Transactions & Missed entirely & Unread leg \\
\midrule
Fluid DEX liquidity layer & 6,428 & 2,999 & 1,262 & 1,737 \\
Curve NG crypto pools & 5,082 & 4,314 & 1,052 & 3,262 \\
Concentrated-liquidity pool (\texttt{Swap(bool,...)}) & 2,998 & 2,668 & 945 & 1,723 \\
DODO & 2,921 & 2,817 & 405 & 2,412 \\
Balancer V1 & 1,885 & 1,622 & 294 & 1,328 \\
Balancer V3 & 1,610 & 1,497 & 199 & 1,298 \\
\textbf{Any of the six} &  & 12,406 & \textbf{3,179} & \textbf{9,227} \\
\bottomrule
\end{tabular*}
\end{minipage}\par\medskip

\subsubsection*{S1.3 Quarterly deployment and strategy participation series}\label{s1.3-quarterly-deployment-and-strategy-participation-series}
\addcontentsline{toc}{subsubsection}{S1.3 Quarterly deployment and strategy participation series}

Tables S1c to S1f report the full quarterly series underlying Section 4.2 and Section 4.3 across top-level contract creations by verified MEV addresses and quarterly participation and concentration in atomic arbitrage, sandwich attacks and lending liquidations.

\par\medskip\noindent\begin{minipage}{\linewidth}
\textbf{Table S1c.} Quarterly top-level contract creations by verified MEV addresses.\par\smallskip
\footnotesize
\renewcommand{\arraystretch}{1.10}
\setlength{\tabcolsep}{2.5pt}
\begin{tabular*}{\linewidth}{@{\extracolsep{\fill}}lrrrr@{}}
\toprule
Quarter & Creations & Deploying addresses & Creations / deployer & Weekend share \\
\midrule
2025Q1 & 3,068 & 69 & 44.46 & 25.62\% \\
2025Q2 & 14,251 & 85 & 167.66 & 31.53\% \\
2025Q3 & 6,640 & 97 & 68.45 & 29.26\% \\
2025Q4 & 3,621 & 98 & 36.95 & 24.88\% \\
2026Q1 & 1,816 & 136 & 13.35 & 30.89\% \\
2026Q2 & 1,516 & 173 & 8.76 & 20.51\% \\
2026Q3 & 1,824 & 241 & 7.57 & 20.23\% \\
\bottomrule
\end{tabular*}
\end{minipage}\par\medskip

\par\medskip\noindent\begin{minipage}{\linewidth}
\textbf{Table S1d.} Quarterly verified atomic arbitrage participation and concentration.\par\smallskip
\scriptsize
\renewcommand{\arraystretch}{1.10}
\setlength{\tabcolsep}{2.5pt}
\begin{tabular*}{\linewidth}{@{\extracolsep{\fill}}lrrrrrr@{}}
\toprule
Quarter & Verified arbitrages & Active addresses & Daily active & Days & Top-50 share & Top-10\% share \\
\midrule
2024Q3 & 114,364 & 621 & 286.2 & 25 & 78.32\% & 81.80\% \\
2024Q4 & 225,318 & 914 & 215.9 & 92 & 80.32\% & 87.85\% \\
2025Q1 & 189,049 & 928 & 220.4 & 90 & 76.81\% & 85.24\% \\
2025Q2 & 232,654 & 849 & 237.3 & 91 & 75.09\% & 83.89\% \\
2025Q3 & 262,314 & 1,073 & 259.8 & 92 & 74.31\% & 86.10\% \\
2025Q4 & 361,334 & 2,077 & 394.4 & 92 & 65.90\% & 88.66\% \\
2026Q1 & 367,905 & 2,357 & 490.8 & 90 & 58.65\% & 84.15\% \\
2026Q2 & 409,488 & 3,011 & 629.4 & 91 & 52.50\% & 80.85\% \\
2026Q3 & 380,873 & 3,749 & 710.5 & 92 & 50.25\% & 82.79\% \\
2026Q4 & 44,320 & 1,855 & 666.0 & 12 & 46.13\% & 71.88\% \\
\bottomrule
\end{tabular*}
\end{minipage}\par\medskip

\par\medskip\noindent\begin{minipage}{\linewidth}
\textbf{Table S1e.} Quarterly verified sandwich participation and concentration.\par\smallskip
\scriptsize
\renewcommand{\arraystretch}{1.10}
\setlength{\tabcolsep}{2.5pt}
\begin{tabular*}{\linewidth}{@{\extracolsep{\fill}}lrrrrrr@{}}
\toprule
Quarter & Verified sandwich transactions & Active addresses & Daily active & Days & Top-50 share & Top-10\% share \\
\midrule
2024Q3 & 201,427 & 223 & 59.4 & 25 & 99.14\% & 97.65\% \\
2024Q4 & 468,934 & 207 & 46.6 & 92 & 99.52\% & 97.32\% \\
2025Q1 & 314,553 & 122 & 51.7 & 90 & 98.81\% & 90.90\% \\
2025Q2 & 398,896 & 118 & 52.9 & 91 & 98.55\% & 89.64\% \\
2025Q3 & 328,664 & 126 & 47.7 & 92 & 98.28\% & 90.58\% \\
2025Q4 & 220,601 & 128 & 45.1 & 92 & 98.55\% & 89.94\% \\
2026Q1 & 214,820 & 111 & 28.1 & 90 & 99.55\% & 93.36\% \\
2026Q2 & 270,629 & 83 & 25.4 & 91 & 99.96\% & 87.62\% \\
2026Q3 & 126,651 & 133 & 40.2 & 92 & 98.14\% & 86.39\% \\
2026Q4 & 13,462 & 83 & 40.1 & 12 & 98.84\% & 83.18\% \\
\bottomrule
\end{tabular*}
\end{minipage}\par\medskip

\par\medskip\noindent\begin{minipage}{\linewidth}
\textbf{Table S1f.} Quarterly verified liquidation participation and concentration.\par\smallskip
\scriptsize
\renewcommand{\arraystretch}{1.10}
\setlength{\tabcolsep}{2.5pt}
\begin{tabular*}{\linewidth}{@{\extracolsep{\fill}}lrrrrrr@{}}
\toprule
Quarter & Verified liquidations & Active addresses & Daily active & Days & Top-50 share & Top-10\% share \\
\midrule
2024Q3 & 540 & 91 & 5.7 & 25 & 91.85\% & 73.33\% \\
2024Q4 & 846 & 98 & 5.5 & 92 & 91.84\% & 63.24\% \\
2025Q1 & 7,511 & 242 & 19.6 & 90 & 83.28\% & 74.10\% \\
2025Q2 & 7,520 & 233 & 12.9 & 91 & 88.15\% & 81.10\% \\
2025Q3 & 2,196 & 105 & 6.8 & 92 & 96.72\% & 82.88\% \\
2025Q4 & 5,399 & 261 & 14.0 & 92 & 83.77\% & 77.44\% \\
2026Q1 & 7,341 & 313 & 18.2 & 90 & 75.83\% & 68.98\% \\
2026Q2 & 5,273 & 327 & 11.7 & 91 & 87.45\% & 83.90\% \\
2026Q3 & 1,516 & 134 & 5.0 & 92 & 92.08\% & 81.27\% \\
2026Q4 & 83 & 39 & 5.2 & 11 & 100.00\% & 40.96\% \\
\bottomrule
\end{tabular*}
\end{minipage}\par\medskip

\subsection*{S2. Liquidation diagnostics and swap execution windows}\label{s2.-liquidation-diagnostics-and-swap-execution-windows}
\addcontentsline{toc}{subsection}{S2. Liquidation diagnostics and swap execution windows}

Table S2 reports the thirty diagnostic specifications for the baseline cloud-screened liquidation panel discussed in Section 4.5 (five event windows on nearby-day and same-weekday calendars in both the full and balanced five-window samples, the shared-reference placement placebo, and the episode strata) under both Monday-anchored weekly clusters and 14-day fortnight clusters (4,000 bootstrap draws).

\par\medskip\noindent\begin{minipage}{\linewidth}
\textbf{Table S2.} Baseline cloud-screened liquidation diagnostics across timing windows, placement placebo and strata.\par\smallskip
\scriptsize
\renewcommand{\arraystretch}{1.10}
\setlength{\tabcolsep}{2.5pt}
\begin{tabular*}{\linewidth}{@{\extracolsep{\fill}}lllrrrll@{}}
\toprule
Calendar or stratum & Support & Window / reference & Events & Obs. / ref. & Ratio & Weekly 95\% CI & Fortnight 95\% CI \\
\midrule
\multicolumn{8}{@{}l@{}}{\textbf{Panel A: Five-window timing profiles}} \\[1pt]
Nearby days & All matched & 6 to 3h before & 81 & 304 / 457.8 & 0.66 & {[}0.24, 1.18{]} & {[}0.24, 1.21{]} \\
 &  & 3 to 1h before & 101 & 232 / 419.6 & 0.55 & {[}0.26, 1.29{]} & {[}0.27, 1.22{]} \\
 &  & Final pre-hour & 102 & 133 / 135.9 & 0.98 & {[}0.24, 2.29{]} & {[}0.22, 2.43{]} \\
 &  & Outage window & 156 & 329 / 1,303.0 & 0.25 & {[}0.11, 0.61{]} & {[}0.11, 0.59{]} \\
 &  & Equal recovery & 80 & 160 / 190.0 & 0.84 & {[}0.35, 1.74{]} & {[}0.35, 1.81{]} \\
 & Balanced 5-window & 6 to 3h before & 29 & 31 / 140.0 & 0.22 & {[}0.08, 0.92{]} & {[}0.09, 1.17{]} \\
 &  & 3 to 1h before & 29 & 104 / 94.0 & 1.11 & {[}0.11, 11.95{]} & {[}0.12, 11.77{]} \\
 &  & Final pre-hour & 29 & 8 / 38.5 & 0.21 & {[}0.07, 0.83{]} & {[}0.07, 1.03{]} \\
 &  & Outage window & 29 & 41 / 55.0 & 0.75 & {[}0.16, 3.04{]} & {[}0.16, 2.98{]} \\
 &  & Equal recovery & 29 & 58 / 50.9 & 1.14 & {[}0.28, 3.46{]} & {[}0.28, 3.54{]} \\
\addlinespace[2.5pt]
Same weekday & All matched & 6 to 3h before & 55 & 129 / 162.8 & 0.79 & {[}0.40, 1.77{]} & {[}0.40, 1.70{]} \\
 &  & 3 to 1h before & 64 & 281 / 313.7 & 0.90 & {[}0.22, 2.36{]} & {[}0.22, 2.42{]} \\
 &  & Final pre-hour & 64 & 90 / 75.8 & 1.19 & {[}0.25, 2.43{]} & {[}0.25, 2.46{]} \\
 &  & Outage window & 84 & 165 / 746.7 & 0.22 & {[}0.10, 0.62{]} & {[}0.10, 0.61{]} \\
 &  & Equal recovery & 45 & 102 / 272.2 & 0.37 & {[}0.15, 1.33{]} & {[}0.14, 1.50{]} \\
 & Balanced 5-window & 6 to 3h before & 12 & 25 / 10.5 & 2.38 & {[}0.23, 11.28{]} & {[}0.24, 13.27{]} \\
 &  & 3 to 1h before & 12 & 24 / 51.3 & 0.47 & {[}0.06, 4.74{]} & {[}0.06, 4.89{]} \\
 &  & Final pre-hour & 12 & 11 / 11.5 & 0.96 & {[}0.26, 6.00{]} & {[}0.26, 6.00{]} \\
 &  & Outage window & 12 & 9 / 16.3 & 0.55 & {[}0.25, 0.83{]} & {[}0.22, 0.81{]} \\
 &  & Equal recovery & 12 & 8 / 110.5 & 0.07 & {[}0.02, 1.17{]} & {[}0.02, 1.36{]} \\
\addlinespace[3pt]
\multicolumn{8}{@{}l@{}}{\textbf{Panel B: Placement placebo and recorded episode strata}} \\[1pt]
Outage day (day 0) & All matched & refs -2 to -5 & 138 & 320 / 1,315.3 & 0.24 & {[}0.11, 0.61{]} & {[}0.11, 0.61{]} \\
 & Balanced & refs -2 to -5 & 75 & 158 / 469.4 & 0.34 & {[}0.23, 0.76{]} & {[}0.23, 0.75{]} \\
\addlinespace[2.5pt]
Pre-outage (day -1) & All matched & refs -2 to -5 & 75 & 328 / 469.4 & 0.70 & {[}0.19, 2.95{]} & {[}0.19, 3.18{]} \\
 & Balanced & refs -2 to -5 & 75 & 328 / 469.4 & 0.70 & {[}0.19, 2.95{]} & {[}0.19, 3.18{]} \\
\addlinespace[2.5pt]
All baseline episodes & Nearby days & Outage window & 156 & 329 / 1,303.0 & 0.25 & {[}0.11, 0.61{]} & {[}0.11, 0.59{]} \\
 & Same weekday & Outage window & 84 & 165 / 746.7 & 0.22 & {[}0.10, 0.62{]} & {[}0.10, 0.61{]} \\
\addlinespace[2.5pt]
Isolated (no overlap) & Nearby days & Outage window & 108 & 205 / 368.0 & 0.56 & {[}0.29, 1.08{]} & {[}0.26, 1.07{]} \\
 & Same weekday & Outage window & 57 & 123 / 302.3 & 0.41 & {[}0.21, 0.91{]} & {[}0.21, 0.89{]} \\
\addlinespace[2.5pt]
API \& coding tools & Nearby days & Outage window & 41 & 117 / 295.1 & 0.40 & {[}0.12, 1.10{]} & {[}0.08, 1.05{]} \\
 & Same weekday & Outage window & 18 & 49 / 216.5 & 0.23 & {[}0.07, 0.44{]} & {[}0.07, 0.44{]} \\
\bottomrule
\end{tabular*}
\end{minipage}\par\medskip

\subsubsection*{S2.1 Paired contrast between the outage window and the placement placebo}\label{s2.1-paired-contrast-between-the-outage-window-and-the-placement-placebo}
\addcontentsline{toc}{subsubsection}{S2.1 Paired contrast between the outage window and the placement placebo}

In the shared-reference placement placebo across 75 balanced episodes, comparing the outage window (day 0) and the clean same-clock window one day earlier (day -1) against the same reference windows two to five days before onset gives a paired outage-to-placebo ratio of 0.48 (weekly 95 percent interval {[}0.17, 1.43{]}; fortnight interval {[}0.16, 1.43{]}) and a paired arm-ratio difference of -0.36 ({[}-2.38, 0.09{]}).

\subsubsection*{S2.2 Verified swap execution across the five outage-panel windows}\label{s2.2-verified-swap-execution-across-the-five-outage-panel-windows}
\addcontentsline{toc}{subsubsection}{S2.2 Verified swap execution across the five outage-panel windows}

Table S2b reports verified atomic arbitrage and sandwich execution across all five event windows in both the baseline cloud-screened and extended samples on both reference calendars (4,000 weekly block-bootstrap draws). Of the 40 window-by-calendar intervals, 36 contain one; below one: sandwich 6 to 3h before (baseline, nearby days) and sandwich 3 to 1h before (baseline, same weekday); above one: arbitrage final pre-hour (baseline, same weekday) and arbitrage outage window (baseline, same weekday).

\par\medskip\noindent\begin{minipage}{\linewidth}
\textbf{Table S2b.} Verified arbitrage and sandwich execution across event windows, samples and calendars.\par\smallskip
\scriptsize
\renewcommand{\arraystretch}{1.10}
\setlength{\tabcolsep}{2.5pt}
\begin{tabular*}{\linewidth}{@{\extracolsep{\fill}}lllrrrrr@{}}
\toprule
Sample & Calendar & Strategy (n) & -6 to -3h & -3 to -1h & -1 to 0h & Outage & Recovery \\
\midrule
Baseline & Nearby days & Arbitrage (156) & 0.99 {[}0.94, 1.05{]} & 0.94 {[}0.88, 1.01{]} & 0.97 {[}0.86, 1.08{]} & 0.95 {[}0.88, 1.04{]} & 1.02 {[}0.94, 1.08{]} \\
 &  & Sandwich (156) & 0.94 {[}0.90, 0.99{]} & 0.94 {[}0.88, 1.00{]} & 0.97 {[}0.91, 1.05{]} & 0.98 {[}0.92, 1.03{]} & 1.00 {[}0.94, 1.08{]} \\
\addlinespace[2.5pt]
Extended & Nearby days & Arbitrage (198) & 0.99 {[}0.93, 1.05{]} & 0.98 {[}0.92, 1.05{]} & 1.00 {[}0.91, 1.09{]} & 0.98 {[}0.91, 1.05{]} & 0.99 {[}0.94, 1.05{]} \\
 &  & Sandwich (198) & 0.96 {[}0.91, 1.00{]} & 0.97 {[}0.92, 1.02{]} & 1.00 {[}0.94, 1.06{]} & 0.97 {[}0.93, 1.02{]} & 1.01 {[}0.95, 1.08{]} \\
\addlinespace[2.5pt]
Baseline & Same weekday & Arbitrage (84) & 1.06 {[}0.95, 1.18{]} & 1.03 {[}0.94, 1.13{]} & 1.15 {[}1.02, 1.26{]} & 1.12 {[}1.03, 1.22{]} & 1.09 {[}0.98, 1.25{]} \\
 &  & Sandwich (84) & 0.98 {[}0.90, 1.09{]} & 0.89 {[}0.82, 0.98{]} & 1.00 {[}0.92, 1.09{]} & 1.00 {[}0.92, 1.09{]} & 1.03 {[}0.92, 1.16{]} \\
\addlinespace[2.5pt]
Extended & Same weekday & Arbitrage (99) & 1.06 {[}0.98, 1.15{]} & 0.97 {[}0.89, 1.07{]} & 1.01 {[}0.90, 1.12{]} & 1.06 {[}0.99, 1.15{]} & 1.06 {[}0.96, 1.17{]} \\
 &  & Sandwich (99) & 0.97 {[}0.89, 1.07{]} & 0.93 {[}0.85, 1.02{]} & 0.99 {[}0.90, 1.08{]} & 1.01 {[}0.93, 1.08{]} & 1.00 {[}0.92, 1.11{]} \\
\bottomrule
\end{tabular*}
\end{minipage}\par\medskip

\subsection*{S3. Cross-chain extensions and the AIDA registry}\label{s3.-cross-chain-extensions-and-the-aida-registry}
\addcontentsline{toc}{subsection}{S3. Cross-chain extensions and the AIDA registry}

\subsubsection*{S3.1 Solana historical strata and archive sizing}\label{s3.1-solana-historical-strata-and-archive-sizing}
\addcontentsline{toc}{subsubsection}{S3.1 Solana historical strata and archive sizing}

To evaluate a non-EVM cross-chain extension, we validated Solana Loader-v3 and Loader-v4 program deployment and upgrade decoding across 2,600 contiguous slot windows and verified Raydium concentrated-liquidity (CLMM) swaps across four 112-slot historical strata bound to three 2024 source eras (Table S3a). Under the transaction-level validation criteria requiring a 41-byte instruction payload, a 205-byte \texttt{SwapEvent} and two child SPL token transfers, 1,010 committed invocations across 951 transactions are verified (536 routed through Jupiter V6), rising to 1,127 swaps when instruction-scoped logs are retained for 13 multi-hop transactions that hit the 10 KB runtime log cap. Because the strata average 45,550.12 decompressed bytes per slot even under two-stage compact queries, a single 90-day discovery window (17,403,249 slots) would require 792.7 GB of archive reads, so we report the verified strata without extrapolating a Solana outage panel.

\par\medskip\noindent\begin{minipage}{\linewidth}
\textbf{Table S3a.} Verified Solana Raydium CLMM swap strata across 2024 source eras.\par\smallskip
\footnotesize
\renewcommand{\arraystretch}{1.10}
\setlength{\tabcolsep}{2.5pt}
\begin{tabular*}{\linewidth}{@{\extracolsep{\fill}}lrrrr@{}}
\toprule
Stratum and UTC date & Slots & CLMM calls (ok / fail) & Strict swaps (txs) & Caller / direct / Jupiter \\
\midrule
Pilot (2024-09-05) & 112 & 1,504 (214 / 1,290) & 180 (170) & 144 / 36 / 105 \\
Early (2024-09-19) & 112 & 1,193 (307 / 886) & 147 (136) & 105 / 42 / 56 \\
Middle (2024-10-20) & 112 & 1,179 (411 / 768) & 334 (321) & 232 / 102 / 177 \\
Late (2024-11-20) & 112 & 2,836 (483 / 2,353) & 349 (324) & 314 / 35 / 198 \\
\textbf{Total (Sep-Nov 2024)} & \textbf{448} & \textbf{6,712 (1,415 / 5,297)} & \textbf{1,010 (951)} & \textbf{795 / 215 / 536} \\
\bottomrule
\end{tabular*}
\end{minipage}\par\medskip

\subsubsection*{S3.2 Full 22-month Base L2 panel and the AIDA multi-vendor AI outage registry}\label{s3.2-full-22-month-base-l2-panel-and-the-aida-multi-vendor-ai-outage-registry}
\addcontentsline{toc}{subsubsection}{S3.2 Full 22-month Base L2 panel and the AIDA multi-vendor AI outage registry}

Table S3b reports per-provider incident counts, primary-source verification counts and developer-relevant exposure hours across all 30 providers in the AI Incident and Disruption Archive (AIDA) from January 2025 through October 2026. Table S3c summarizes the hourly within-day PPML and daily lag-2 dose-response estimates across the 15,600-hour target period on both Ethereum L1 and Base L2 (1,099,593,191 Base swaps and 29,120 private executors across 21,026 operator addresses).

\par\medskip\noindent\begin{minipage}{\linewidth}
\textbf{Table S3b.} Per-provider coverage of the AIDA multi-vendor outage registry (January 2025 to October 2026).\par\smallskip
\footnotesize
\renewcommand{\arraystretch}{1.10}
\setlength{\tabcolsep}{2.5pt}
\begin{tabular*}{\linewidth}{@{\extracolsep{\fill}}llrrr@{}}
\toprule
Provider or group & Region and category & Raw rows & Deduplicated (verified) & Dev 60m-24h (hours) \\
\midrule
\multicolumn{5}{@{}l@{}}{\textbf{Panel A: Proprietary foundation-model vendors (17 providers)}} \\[1pt]
Anthropic (Claude API, Console, Code) & International model API & 1,529 & 727 (707, 97\%) & 303 (838.2h) \\
\addlinespace[2.5pt]
OpenAI (API, Codex, Responses, Batch) & International model API & 1,299 & 650 (628, 97\%) & 175 (887.0h) \\
\addlinespace[2.5pt]
Moonshot Kimi & Chinese model API & 869 & 579 (577, 100\%) & 60 (180.2h) \\
\addlinespace[2.5pt]
MiniMax & Chinese model API & 494 & 412 (374, 91\%) & 5 (27.3h) \\
\addlinespace[2.5pt]
Mistral (La Plateforme) & International model API & 237 & 201 (188, 94\%) & 15 (30.9h) \\
\addlinespace[2.5pt]
xAI (Grok API) & International model API & 290 & 196 (171, 87\%) & 61 (234.4h) \\
\addlinespace[2.5pt]
DeepSeek & Chinese model API & 193 & 104 (95, 91\%) & 20 (47.7h) \\
\addlinespace[2.5pt]
Google AI (Gemini API, Vertex) & International model API & 163 & 97 (79, 81\%) & 37 (217.7h) \\
\addlinespace[2.5pt]
Perplexity (Sonar API) & International model API & 79 & 33 (32, 97\%) & 6 (34.7h) \\
\addlinespace[2.5pt]
Cohere & International model API & 52 & 28 (28, 100\%) & 15 (55.0h) \\
\addlinespace[2.5pt]
Zhipu BigModel (GLM) & Chinese model API & 28 & 22 (17, 77\%) & 3 (9.8h) \\
\addlinespace[2.5pt]
Volcengine Doubao & Chinese model API & 11 & 9 (7, 78\%) & 1 (1.1h) \\
\addlinespace[2.5pt]
Baidu Qianfan (ERNIE) & Chinese model API & 15 & 7 (7, 100\%) & 0 (0.0h) \\
\addlinespace[2.5pt]
Alibaba Qwen (DashScope) & Chinese model API & 7 & 4 (3, 75\%) & 1 (5.2h) \\
\addlinespace[2.5pt]
Tencent Hunyuan & Chinese model API & 10 & 4 (4, 100\%) & 4 (11.9h) \\
\addlinespace[2.5pt]
Meta Llama API & International model API & 4 & 2 (2, 100\%) & 1 (2.6h) \\
\addlinespace[2.5pt]
Other Chinese vendors (6 platforms) & Chinese model API & 11 & 10 (4, 40\%) & 2 (2.8h) \\
\addlinespace[3pt]
\multicolumn{5}{@{}l@{}}{\textbf{Panel B: Cloud AI hosts, coding environments and infrastructure controls (13 providers)}} \\[1pt]
Azure OpenAI \& AWS Bedrock & Managed cloud AI & 67 & 41 (22, 54\%) & 0 (0.0h) \\
\addlinespace[2.5pt]
Cursor \& GitHub & Coding \& platform & 310 & 302 (32 verified) & 270 controls \\
\addlinespace[2.5pt]
Cloudflare, QuickNode, Alchemy, Infura & RPC \& edge infra & 2,332 & 2,293 (statuspage) & 2,293 controls \\
\addlinespace[2.5pt]
AWS, npm, GCP, Azure, PyPI & Cloud \& pkg infra & 189 & 147 (statuspage) & 147 controls \\
\addlinespace[2.5pt]
\textbf{Total AIDA registry (30 providers)} & \textbf{All 4 layers} & \textbf{8,189} & \textbf{5,868 (2,977 verified)} & \textbf{709 (2,586.5h)} \\
\bottomrule
\end{tabular*}
\end{minipage}\par\medskip

\par\medskip\noindent\begin{minipage}{\linewidth}
\textbf{Table S3c.} Cross-chain deployment and execution responses on Ethereum and Base (15,600 UTC hours).\par\smallskip
\scriptsize
\renewcommand{\arraystretch}{1.10}
\setlength{\tabcolsep}{2.5pt}
\begin{tabular*}{\linewidth}{@{\extracolsep{\fill}}llrrrr@{}}
\toprule
Margin & Outcome & Total count & All vendors & OpenAI hours & Daily lag 2 (8h) \\
\midrule
\multicolumn{6}{@{}l@{}}{\textbf{Panel A: Ethereum L1 (22 months, 15,600 UTC hours)}} \\[1pt]
Development & Private executor creations & 2,709 & 0.96 {[}0.67, 1.38{]} & 0.68 {[}0.48, 0.94{]} & 0.66 {[}0.51, 0.86{]} \\
 & Distinct addresses creating executor & 2,366 & 0.91 {[}0.69, 1.21{]} & 0.69 {[}0.49, 0.97{]} & 0.66 {[}0.51, 0.86{]} \\
\addlinespace[2.5pt]
Execution & Verified arbitrage execution & 4,947,869 & 0.99 {[}0.96, 1.02{]} & 1.02 {[}0.98, 1.06{]} & 1.00 {[}0.98, 1.02{]} \\
 & All pool swap transactions & 135,465,042 & 0.99 {[}0.98, 1.00{]} & 1.00 {[}0.98, 1.02{]} & 1.00 {[}0.98, 1.02{]} \\
\addlinespace[3pt]
\multicolumn{6}{@{}l@{}}{\textbf{Panel B: Base L2 (22 months, 15,600 UTC hours)}} \\[1pt]
Development & Operator top-level creations (signed) & 93,694 & 0.94 {[}0.82, 1.09{]} & 0.76 {[}0.59, 0.99{]} & 1.04 {[}0.91, 1.19{]} \\
 & Distinct operators deploying executor & 15,440 & 1.00 {[}0.92, 1.09{]} & 0.98 {[}0.84, 1.14{]} & 0.88 {[}0.78, 0.98{]} \\
 & Executor creations (capped <=3/op/h) & 19,459 & 1.00 {[}0.92, 1.09{]} & 0.97 {[}0.83, 1.13{]} & 0.89 {[}0.78, 1.00{]} \\
 & Newly appearing operator caller keys & 21,026 & 1.04 {[}0.93, 1.16{]} & 1.20 {[}1.00, 1.44{]} & 0.88 {[}0.78, 0.99{]} \\
\addlinespace[2.5pt]
Execution & Arbitrage executor transactions & 26,165,900 & 1.00 {[}0.97, 1.02{]} & 0.95 {[}0.91, 0.99{]} & 1.00 {[}0.97, 1.04{]} \\
 & All pool swap transactions & 1,099,593,191 & 1.00 {[}0.98, 1.01{]} & 0.99 {[}0.97, 1.02{]} & 1.01 {[}0.98, 1.03{]} \\
\bottomrule
\end{tabular*}
\par\smallskip{\footnotesize Notes: All-vendor and OpenAI-hour columns report hourly PPML rate ratios with UTC-date and hour-of-day fixed effects and UTC-date-clustered 95\% CIs; final column reports daily PPML rate ratios two days after an 8h Anthropic or OpenAI disruption.}
\end{minipage}\par\medskip

\subsection*{S4. Supplementary AI quota and access announcements}\label{s4.-supplementary-ai-quota-and-access-announcements}
\addcontentsline{toc}{subsection}{S4. Supplementary AI quota and access announcements}

Table S4 reports cohort sizes, window days, top-level contract creations and matched dates for the six 2026 AI access and quota announcements plotted in Figure 7.

\par\medskip\noindent\begin{minipage}{\linewidth}
\textbf{Table S4.} Daily contract creations around 2026 AI access and quota announcements.\par\smallskip
\scriptsize
\renewcommand{\arraystretch}{1.10}
\setlength{\tabcolsep}{2.5pt}
\begin{tabular*}{\linewidth}{@{\extracolsep{\fill}}lllrrrr@{}}
\toprule
Provider & Announcement date & Mechanism & Cohort / days & Creation txs & Creating-sender days & Matched dates \\
\midrule
OpenAI & 2 Feb 2026 & Policy expansion & 260 / 22 & 19 & 15 & 3 (Feb 1, 8, 15) \\
 & 28 Apr 2026 & Broad replenishment & 312 / 22 & 61 & 35 & 0 \\
 & 11 Jun 2026 & Banked grant & 148 / 22 & 220 & 50 & 0 \\
 & 8/13 Aug 2026 & Broad replenishment & 324 / 27 & 92 & 44 & 0 \\
\addlinespace[2.5pt]
Anthropic & 6 May 2026 & Policy expansion & 312 / 22 & 56 & 26 & 0 \\
 & 22 Sep 2026 & Banked grant & 314 / 22 & 70 & 28 & 0 \\
\bottomrule
\end{tabular*}
\end{minipage}\par\medskip

\subsection*{S5. Cycle-search frontier in dense router transactions}\label{s5.-cycle-search-frontier-in-dense-router-transactions}
\addcontentsline{toc}{subsection}{S5. Cycle-search frontier in dense router transactions}

The arbitrage detector enumerates simple token cycles inside each transaction under a 100,000-state search cap after decomposing the transaction\textquotesingle s token graph into strongly connected components to prune dead-end edges. When a dense multi-pool aggregator transaction contains parallel swap legs across many pools within a single strongly connected component whose concrete leg combinations exceed the 100,000-state cap, the transaction is recorded as unresolved, receives neither a positive nor a negative arbitrage label, and remains eligible for sandwich matching. Across the full Ethereum ledger, 42 transactions (0.25 per million observed transactions) reach this cap (Table S1a).

\subsection*{S6. Cloud-status curation and the baseline cloud-screened sample}\label{s6.-cloud-status-curation-and-the-baseline-cloud-screened-sample}
\addcontentsline{toc}{subsection}{S6. Cloud-status curation and the baseline cloud-screened sample}

The public cloud-status screening table matched 16 cloud-disruption records to study windows. Twelve carry verified closing timestamps, whereas four regional cloud-status records beginning in late April 2026 end at ongoing-disruption or partial-recovery updates without a verified resolution timestamp. The baseline cloud-screened sample retains the 12 explicit-recovery records and excludes windows after the onset of the unresolved regional records (156 nearby-day, 84 same-weekday and 64 balanced liquidation episodes), whereas the extended full-period sample retains all 198 nearby-day, 98 same-weekday and 73 balanced episodes through October 2026. On the episodes present in both samples, point estimates and confidence intervals are identical.

\subsection*{S7. Creation windows for verified addresses and prospective liquidation cohorts}\label{s7.-creation-windows-for-verified-addresses-and-prospective-liquidation-cohorts}
\addcontentsline{toc}{subsection}{S7. Creation windows for verified addresses and prospective liquidation cohorts}

Alongside the chain-wide typed creation panel of Section 4.4, we also evaluated episode-window top-level contract creations in two narrower sets. Across all 8,516 verified MEV addresses (33,066 top-level creations over the target period), raw outage-window creation ratios sit near one (0.85, {[}0.68, 1.09{]} in the baseline nearby-day sample and 0.85, {[}0.69, 1.07{]} in the extended sample) because untyped creations mix trading executors with non-trading contracts and single-use constructor-executed trades. Within the narrower prospective liquidation cohorts (733 qualifying addresses and 3,398 top-level creations by 181 addresses), episode-window creation counts are too sparse to support inference on their own, as the 64 balanced baseline episodes contain only a single creation (Table S7).

\par\medskip\noindent\begin{minipage}{\linewidth}
\textbf{Table S7.} Contract-creation and creating-sender responses for prospective liquidation cohorts.\par\smallskip
\scriptsize
\renewcommand{\arraystretch}{1.10}
\setlength{\tabcolsep}{2.5pt}
\begin{tabular*}{\linewidth}{@{\extracolsep{\fill}}llrrrr@{}}
\toprule
Sample & Reference calendar & Observed / reference & Rate ratio & 95\% interval & n \\
\midrule
\multicolumn{6}{@{}l@{}}{\textbf{Panel A: Top-level contract creations}} \\[1pt]
Extended & Nearby days, all & 51 / 31.60 & 1.61 & {[}0.84, 2.71{]} & 198 \\
 & Same weekday, all & 9 / 9.00 & 1.00 & {[}0.43, 2.00{]} & 98 \\
 & Nearby days, balanced & 6 / 8.23 & 0.73 & {[}0.30, 1.24{]} & 73 \\
 & Same weekday, balanced & 6 / 6.67 & 0.90 & {[}0.24, 1.85{]} & 73 \\
\addlinespace[2.5pt]
Baseline & Nearby days, all & 15 / 9.50 & 1.58 & {[}0.32, 3.74{]} & 156 \\
 & Same weekday, all & 3 / 5.67 & 0.53 & {[}0.00, 1.60{]} & 84 \\
 & Nearby days, balanced & 1 / 3.25 & 0.31 & {[}0.00, 1.09{]} & 64 \\
 & Same weekday, balanced & 1 / 4.83 & 0.21 & {[}0.00, 0.75{]} & 64 \\
\addlinespace[3pt]
\multicolumn{6}{@{}l@{}}{\textbf{Panel B: Creating sender addresses}} \\[1pt]
Extended & Nearby days, all & 24 / 19.90 & 1.21 & {[}0.63, 2.11{]} & 198 \\
\addlinespace[2.5pt]
Baseline & Nearby days, all & 6 / 8.00 & 0.75 & {[}0.21, 1.39{]} & 156 \\
 & Nearby days, balanced & 1 / 2.92 & 0.34 & {[}0.00, 1.20{]} & 64 \\
 & Same weekday, balanced & 1 / 4.33 & 0.23 & {[}0.00, 0.78{]} & 64 \\
\bottomrule
\end{tabular*}
\par\smallskip{\footnotesize Notes: Prospective 90-day liquidation cohorts; 4,000 Monday-week bootstrap draws.}
\end{minipage}\par\medskip

\end{document}